\documentclass[10pt,preprint2]{aastex}
\usepackage{color}
\usepackage{ulem}
\usepackage{multirow}

\newcommand{\figref}[1]{Fig.~\ref{#1}}
\newcommand{\eqref}[1]{eq.~(\ref{#1})}
\newcommand{\tab}[1]{Tab.~(\ref{#1})}
\newcommand{\etal}{\textit{et al.}}

\newcommand{\indice}[1]{_{\textrm{\tiny #1}}}

\shorttitle{H-He-$Z$ mixtures}
\shortauthors{Soubiran \& Militzer}

\begin{document}

\title{The properties of heavy elements in giant planet envelopes}

\author{Fran\c{c}ois Soubiran\altaffilmark{1} and Burkhard
Militzer\altaffilmark{1,2}}
\affil{$^1$Department of Earth and Planetary Science, University of California,
    Berkeley, CA 94720, United States}
\affil{$^2$Department of Astronomy, University of California,
    Berkeley, CA 94720, United States}

\begin{abstract}
  The core accretion model for giant planet formation suggests a two layer picture 
for the initial structure of Jovian planets, with heavy elements in a dense core 
and a thick H-He envelope. Late planetesimal accretion and core erosion could 
potentially enrich the H-He envelope in heavy elements, which is supported by the 
three-fold solar metallicity that was measured in Jupiter's atmosphere by the 
Galileo entry probe. In order to reproduce the observed gravitational moments of 
Jupiter and Saturn, models for their interiors include heavy elements, $Z$, in 
various proportions. However, their effect on the equation of state of the 
hydrogen-helium mixtures has not been investigated beyond the ideal mixing 
approximation. In this article, we report results from \textit{ab initio} 
simulations of fully interacting H-He-$Z$ mixtures in order to characterize their 
equation of state and to analyze possible consequences for the interior 
structure and evolution of giant planets. Considering C, N, O, Si, Fe, MgO and 
SiO$_2$, we show that the behavior of heavy elements in H-He mixtures may still be 
represented by an ideal mixture if the effective volumes and internal energies are 
chosen appropriately. In the case of oxygen, we also compute the effect on the 
entropy. We find the resulting changes in the temperature-pressure profile to be 
small. A homogeneous distribution of 2\% oxygen by mass changes the temperature in 
Jupiter's interior by only 80 K.  

\date{\today}

\end{abstract}


\section{Introduction}
Despite numerous observations of giant planets in our solar system 
\citep{bolton_2010,jones_2015} and the extrasolar campaigns 
\citep{wright_2011,schneider_2011}, our understanding of the structure, evolution, 
and formation of Jovian planets remains uncertain \citep{guillot_2015}. Largely, 
the uncertainty is due to the fact that the observations of Jovian planets only 
provide data on global properties, which limits the constraints that can be placed 
on interior properties. While future observations, such as the Juno mission, will 
provide more detailed data, additional constraints on Jovian interiors can also be 
gleaned from advances in numerical simulations \citep{militzer_2013} and 
experiments \citep{brygoo_2015} on high density matter and the subsequent 
improvements in planetary models \citep{nettelmann_2012,helled_2013,hubbard_2016}.  

Even in the solar system, the exact composition of the gaseous giant planets is 
still not very well constrained because the composition of the observable 
atmosphere is not necessarily representative of the entire planet. Following the 
core-accretion hypothesis  \citep{pollack_1996}, Jupiter and Saturn were formed by 
a rapid accretion of solid material until a critical mass of approximately 
10~$M_\Earth$ had been reached, which triggers a substantial gas accretion. Once 
the envelope is as massive as the core, it even becomes a run-away accretion that 
stops when all the gas in the nebula has disappeared, after about 10 Myr. This 
minimum mass of 10~$M_\Earth$ provides also a lower bound for the average 
metallicity of Jupiter and Saturn. As the total mass of Jupiter (resp. Saturn) is 
317.8~$M_\Earth$ (resp. 95.1~$M_\Earth$) \citep{baraffe_2010}, the minimal 
metallicity is $Z\indice{J}=0.031$ (resp. $Z\indice{S}=0.105$) which is higher 
than the solar value of $Z_\Sun=0.0149$ \citep{lodders_2003}. But this is only a 
minimum value, and for Jupiter, the Galileo entry probe measured an atmospheric 
metallicity of 3 times the solar value \citep{wong_2004} for instance.

The dichotomy between a dense core and a H-He envelope is however, at best, only 
an approximated view of the giant planets as there are at least five different 
reasons why the heavy element distribution in giant planets is uncertain. The 
first one comes from the formation of the planet. During the gas accretion, the 
planet kept accreting additional planetesimals and it is unclear whether they 
dissolved in the envelope or reached the core \citep{fortney_2013}.  

The second reason is the possible erosion of the core as proposed by 
\citet{stevenson_1982}, which could trigger a redistribution of the heavy elements 
of the core throughout the envelope. This possibility has been recently 
investigated with \textit{ab initio} simulations, which predicted that all the 
dominant species in the core are miscible in metallic hydrogen 
\citep{WilsonMilitzer2012,WilsonMilitzer2012b,wahl_2013,gonzalez_2014}. 

However, and this is the third reason, the kinetics of the erosion process is very 
poorly constrained and could be very slow. Once dissolved in hydrogen, the heavy 
elements may not be efficiently redistributed throughout the envelope because the 
density contrast may be too high for the convection to advect these elements 
against the gravitational forces. Thus, a semi-convective pattern is likely to set 
in, inducing a gradient of composition in the layers close to the core 
\citep{leconte_2012,leconte_2013}.

An other origin of heterogeneity in giant planet envelopes could come from the 
phase separation of H-He which has been proposed as an explanation of the excess 
of luminosity of Saturn \citep{stevenson_1977,fortney_2004}. The exact phase 
diagram is still controversial despite recent work using \textit{ab initio} 
simulations  \citep{lorenzen_2009,lorenzen_2011} and even thermodynamic 
integrations to properly account for the non-ideal entropy 
\citep{morales_2009,morales_2013}. However, some experimental constraints should 
be available shortly using reflectivity measurements in laser-driven shock 
experiments \citep{soubiran_2013}. But if the phase separation occurs in a giant 
planet, it could also inhibit the advection of heavy elements from the deep 
interior to the external layers as the convection would most likely be rendered 
less efficient by the phase separation. 

The last reason of uncertainty comes from the possible partitioning of the heavy 
elements due to the miscibility difference with helium and with hydrogen. It has 
been suggested, for instance, as an explanation for the strong depletion in neon 
in the atmosphere of Jupiter \citep{WilsonMilitzer2010}.

We therefore need to have better constraints on the distribution of the heavy 
elements. They may come from both observations and more detailed models. 
Measurements of the gravitational field of Jupiter and Saturn via the study of the 
trajectory of their satellites but also via direct measurements from spacecrafts 
such as Cassini and Juno can give strong constraints on the gravitational moments 
of the planet and thus on the mass distribution. A new field is also emerging: 
seismology of giant planets. Recent results using planetary oscillations 
\citep{gaulme_2011} and more importantly through ring seismology 
\citep{fuller_2014a,fuller_2014b} have started to demonstrate its potential and 
will most likely lead to additional constraints.

\begin{figure}[!t]
\centering
\includegraphics[width=\columnwidth]{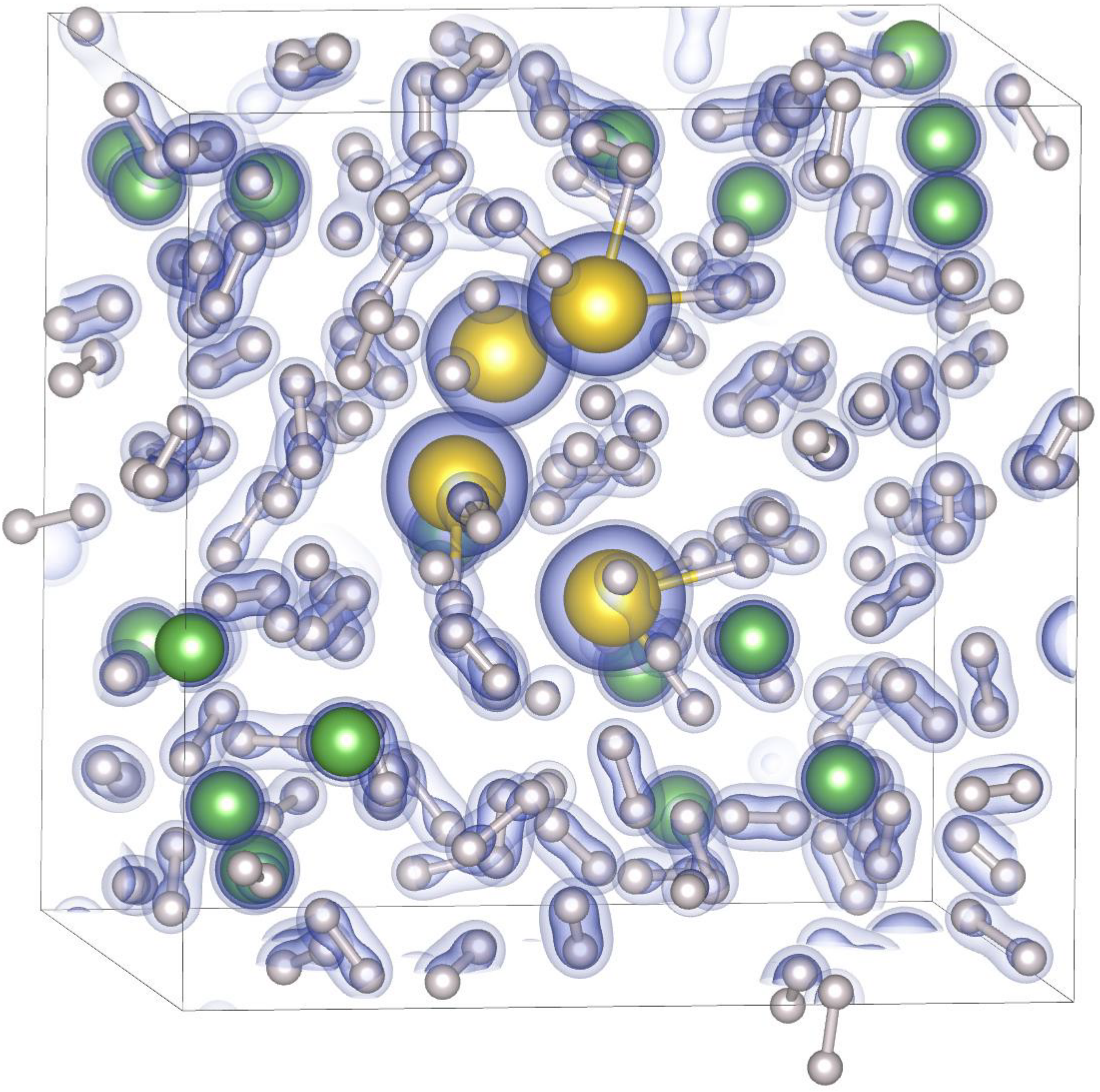}
\caption{\label{fig:simulation_box} Snapshot of a simulation box containing 220 H 
(white), 18 He (green) and 4 Fe (yellow) atoms at 2000~K and 60~GPa. We used a 
distance analysis to identify the chemical bonds. The isosurfaces represent the 
electronic density. }
\end{figure}

Theorists can also help to constrain the possible structures by building detailed 
models including accurate physical properties of matter at high density. 
Improvements in \textit{ab initio} simulations over the past two decades offered 
the possibility to compute precise equations of state (EOS). Recently an updated 
EOS of H-He has been developed by \citet{militzer_2013}. Concerning the heavy 
elements, different EOSs have been used in giant planet models (see 
\citet{baraffe_2008} for a review of the common EOSs). Recent models showed that 
the heavy elements altered the density profile and therefore the structure of the 
planet \citep{hubbard_2016}, as well as its evolution \citep{baraffe_2008}. 
However, these models rely on the assumption that the multi-component mixtures of 
interest are ideal mixtures with an isothermal-isobaric additive mixing rule. 
Under this assumption, one derives all extensive thermodynamic properties of a 
mixture -- volume, internal energy,... -- by adding up the contributions from the 
individual pure species at a given temperature and pressure. Such a mixing rule 
neglects all the inter-species interactions, although they may be much more 
important than the intra-species ones. For instance, in the diluted limit, a heavy 
atom only interacts with the H-He -- solvent -- atoms while, in the ideal mixing 
approximation, the properties of the heavy species are taken from a system of 
heavy atoms only.

Once the Juno mission will have measured the gravitational moments of Jupiter with 
high precision, interior models with various amounts of heavy elements in the 
envelope and different core sizes will be constructed to match these measurements. 
This will improve our understanding of the interior structure and the evolution of 
Jupiter. One goal of this paper is to make this analysis more accurate by going 
beyond the standard ideal mixing rule and by properly characterizing the influence 
of heavy elements on a H-He envelope.  

In this article, we investigate the thermodynamic properties of ternary mixtures 
of hydrogen, helium and heavy elements -- namely carbon, nitrogen, oxygen, 
silicon, iron as well as magnesium oxide and silicon dioxide -- under conditions 
relevant for the giant planet interiors. We used \textit{ab initio} simulations to 
determine the influence of these seven elements on the density and the internal 
energy. On the case of oxygen, we also studied the influence on the entropy, which 
is essential to determine the pressure-temperature profile in giant planet 
envelopes.

From our analysis we determined that the ternary mixtures can indeed be very well 
described by an ideal isothermal-isobaric mixing of H-He on one side and the heavy 
element on the other, provided however that effective volume or energy of the 
heavy species are chosen appropriately. Both properties may differ from the 
properties from those of the pure species at the same temperature and conditions. 
They may furthermore be affected by the dissociation of hydrogen.

We also performed entropy calculations for H-He-O ternary mixtures. We show that 
the addition of heavy element, homogeneously throughout the envelope of a giant 
planet, has only a marginal influence on the pressure-temperature and on the 
density profiles. Last, we explore the influence of the heavy elements on the 
mass-radius relationship of giant planets.

\section{Simulation methods}
We performed molecular dynamics (MD) simulations with forces derived from density 
functional theory (DFT) \citep{hohenberg_1964} to treat the quantum behavior of 
the electrons. We performed the simulations with the Vienna \textit{ab initio} 
simulation package \citep{kresse_1996}. We used cubic cells with periodic boundary 
conditions. Starting from cells with 220 hydrogen and 18 helium atoms 
\citep{militzer_2013}, we added from 2 to 8 atoms or molecules of heavy elements 
-- we considered C, N, O, Si, Fe, MgO and SiO$_2$. As an example, 
\figref{fig:simulation_box} shows a representative snapshot of a simulation of H, He 
and Fe.

\begin{figure}[!t]
\centering
\includegraphics[width=\columnwidth]{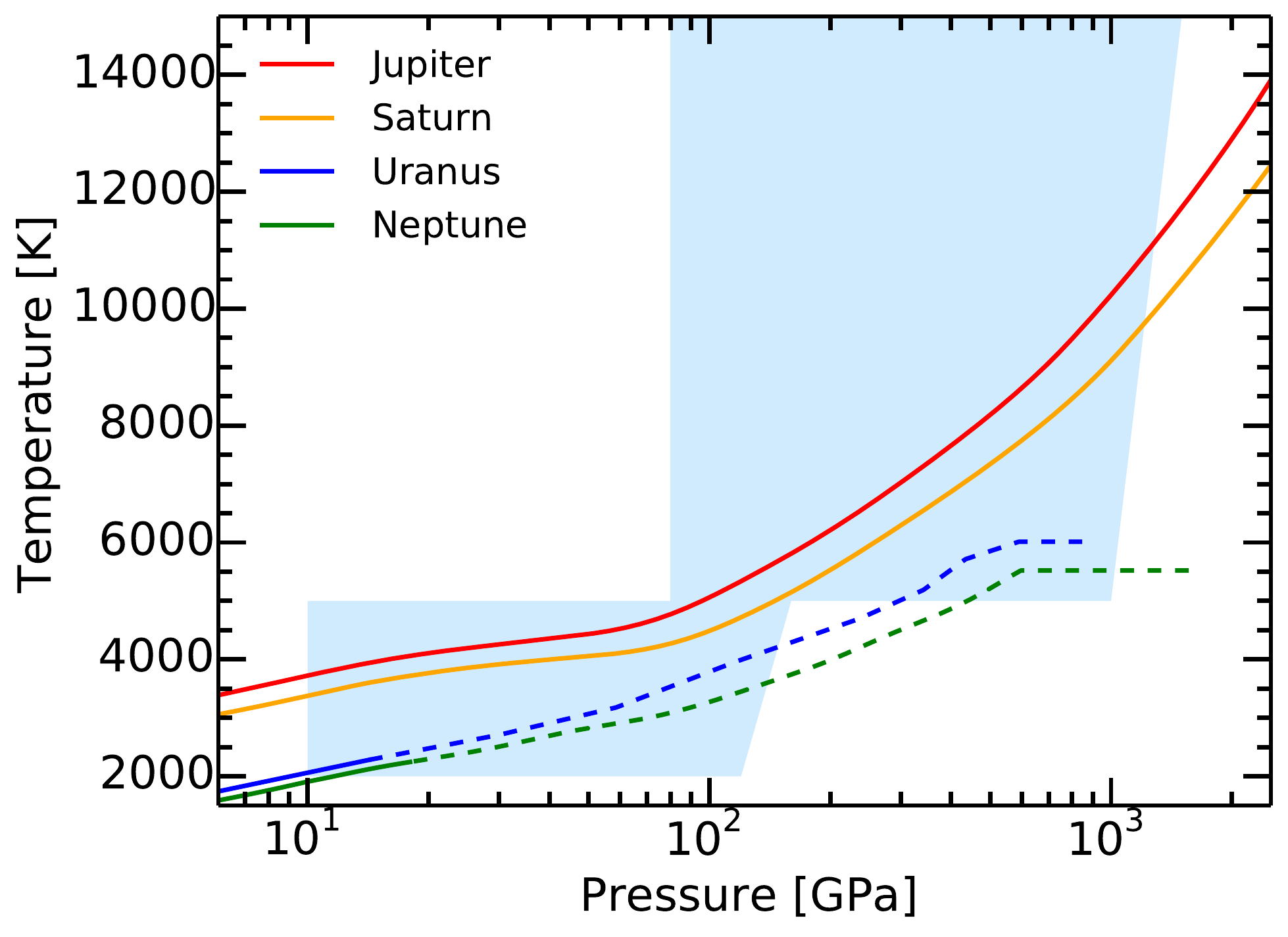}
\caption{\label{fig:P-T_diagram} Pressure vs temperature profiles of the solar 
giant planets. Jupiter and Saturn adiabatic profiles are from 
\citet{militzer_2013b} and Uranus and Neptune models are from 
\citet{nettelmann_2013}. The dashed lines indicate the part of the planets that 
is expected to be mostly made of heavy elements. The shaded region represents 
the range of parameters we explored.}
\end{figure}

We performed the dynamics with a 0.2~fs time step, for a trajectory of at least 
1~ps. The temperature was controlled by a Nos\'e thermostat 
\citep{nose_1984,nose_1991}. We solved the DFT part using the Kohn-Sham scheme 
\citep{kohn_1965} at finite temperature \citep{mermin_1965} with a Fermi-Dirac 
distribution function. We employed Perdew, Burke and Ernzerhof (PBE) 
\citep{perdew_1996} exchange-correlation functionals. This functional provided 
accurate results for numerous systems including the pure species presently 
considered \citep{caillabet_2011, militzer_2009, 
benedict_2014,driver_2015,driver_2016,militzer_2015,denoeud_2014}. We employed 
standard VASP projector augmented wave (PAW) pseudo-potentials 
\citep{blochl_1994}. The cutoff radius was 0.8~$a_0$ for hydrogen, 1.1~$a_0$ for 
helium, 1.1~$a_0$ for carbon, nitrogen and oxygen, each with a 1s$^2$ frozen core, 
2.0~$a_0$ for magnesium with a 1s$^2$2s$^2$ core, 1.9~$a_0$ for silicon with a 
1s$^2$2s$^2$2p$^6$ core, and  2.2~$a_0$ for iron with a 
1s$^2$2s$^2$2p$^6$3s$^2$3p$^6$ core, where $a_0$ stands for the Bohr radius. We 
used the frozen core approximation to speed-up the calculations and because we are 
at low enough density for the core energy level shift towards the continuum to be 
negligible, as well as at low enough temperatures for the thermal ionization of 
the core levels to be insignificant \citep{driver_2015}. The energy cutoff for the 
plane-wave basis was set to 1200~eV. The number of electronic bands was adapted to 
the species, the density and the temperature conditions in order to completely 
cover the spectrum of the fully and partially occupied eigenstates. We sampled the 
Brillouin zone with the Baldereschi point \citep{baldereschi_1973}. 
\citet{militzer_2013} showed that the convergence of the calculation for H-He was 
very good with this choice of parameters.

On a subset of simulations, we also performed thermodynamic integrations to obtain 
the Helmholtz free energy and the entropy. We progressively switched the 
interactions in the system from the potential given by the DFT $U\indice{DFT}$ to 
a classical potential $U\indice{cl}$. This methods has been applied on many 
systems \citep{wijs_1998, 
morales_2009,WilsonMilitzer2010,WilsonMilitzer2012,WilsonMilitzer2012b, 
McMahon2012,wahl_2013,wahl_2015}. It provides the Helmotz free energy difference 
between the two systems: 
\begin{equation}
 F\indice{DFT} - F\indice{cl} = \int_0^1 \textrm{d}\lambda\,\langle 
U\indice{DFT}-U\indice{cl} \rangle_\lambda ,
\end{equation}
where the parameter $\lambda$ defines the hybrid potential 
$U_\lambda=U\indice{cl}+\lambda (U\indice{DFT} - U\indice{cl})$ and the average 
refers to an average over configurations computed with the potential $U_\lambda$. 
For the classical potential, we used a set of non-bonding pair potentials fitted 
on the DFT forces. More details of the integration are given in 
\citet{soubiran_2015} and in \citet{wahl_2015}.

\begin{figure}[t!]
\centering
\includegraphics[width=\columnwidth]{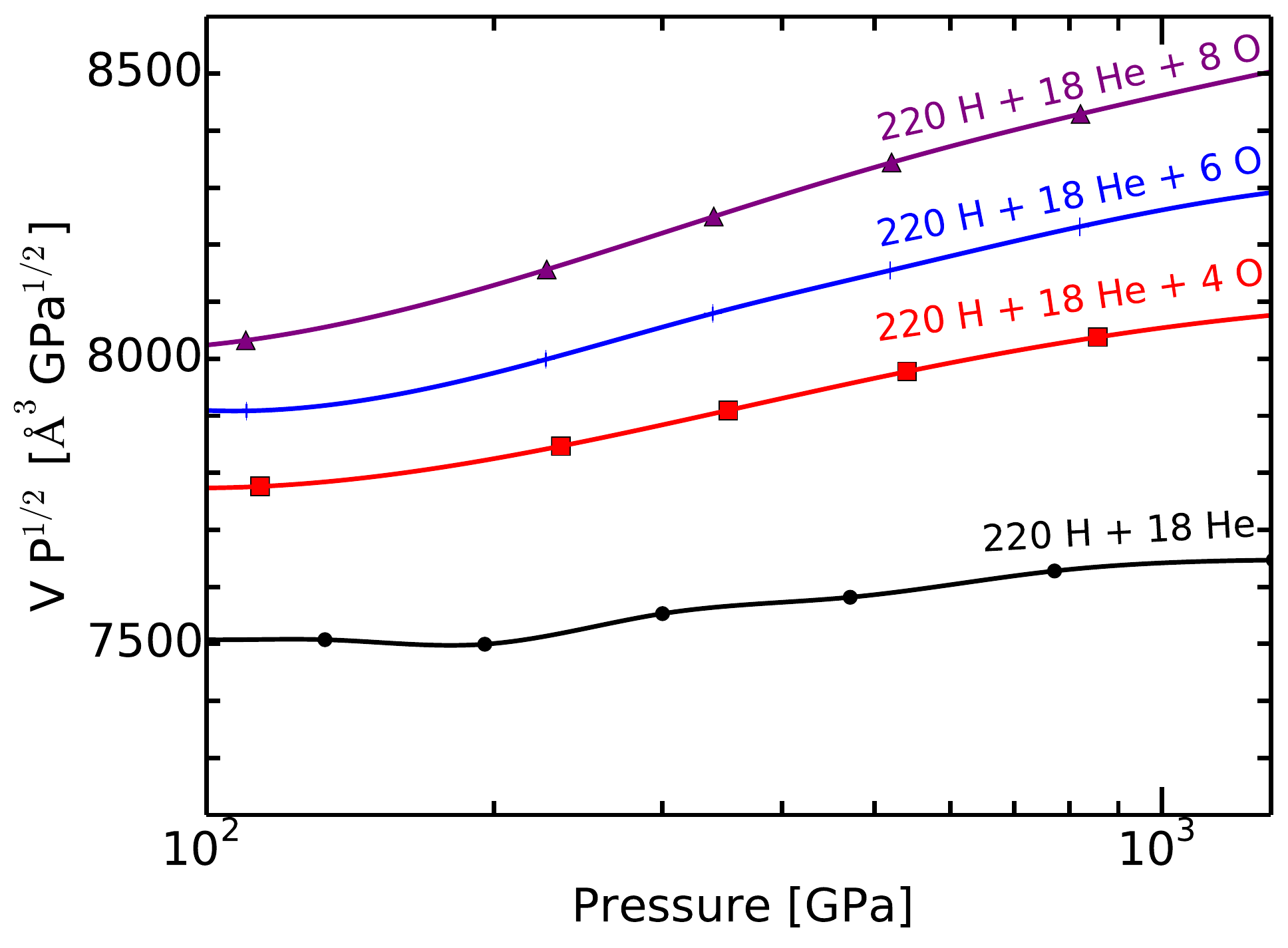}
\caption{\label{fig:volumetot} Volume of the H-He-O simulation cell multiplied by 
the square root of the pressure -- to magnify the differences -- as a function of 
the pressure at 10000~K for different oxygen composition. The symbols show the 
numerical results and the curves represent spline interpolations.}
\end{figure}

In order to determine the different thermodynamic quantities as a function of the 
pressure rather than the density, we used a spline interpolation method. For the 
pressure and the volume, we used the logarithmic values as variables for the 
interpolation. \figref{fig:volumetot} shows an example of an interpolation.  

\section{Results}
We investigated the properties of ternary mixtures with H-He and seven different 
heavy elements: C, N, O, Si, Fe, MgO, SiO$_2$. We explored the thermodynamic 
properties on two different domains: from 10 to 150 ~GPa and 2000 to 5000~K, and 
from 80 to 1500~GPa and 5000 to 15000~K -- see \figref{fig:P-T_diagram}. The 
thermodynamic data are presented in Tabs. \ref{table:eosH-He-Z} and 
\ref{table:eosH-He-O}. We also added the numerical values of the effective 
properties we computed in Tab.~\ref{table:effprop}. For each simulation we 
computed a one-$\sigma$ errorbar for the thermodynamic data, based on a block 
averaging method \citep{rapaportbook}. We used standard error propagation methods 
for the effective properties.

\subsection{Effective volume \label{sec:effvol}}
For each heavy element, we performed MD-DFT simulations for 3 different 
concentrations, keeping the number of hydrogen and helium atoms constant, 
H:He=220:18. We deliberately used a small number of heavy element atoms -- from 2 
to 12 -- in order to stay in the diluted limit where the interaction effects 
between the heavy species are small. Results for the ternary H-He-$Z$ mixture were 
systematically compared with the binary H-He mixture from \citet{militzer_2013}.

\begin{figure}[!t]
\centering
\includegraphics[width=\columnwidth]{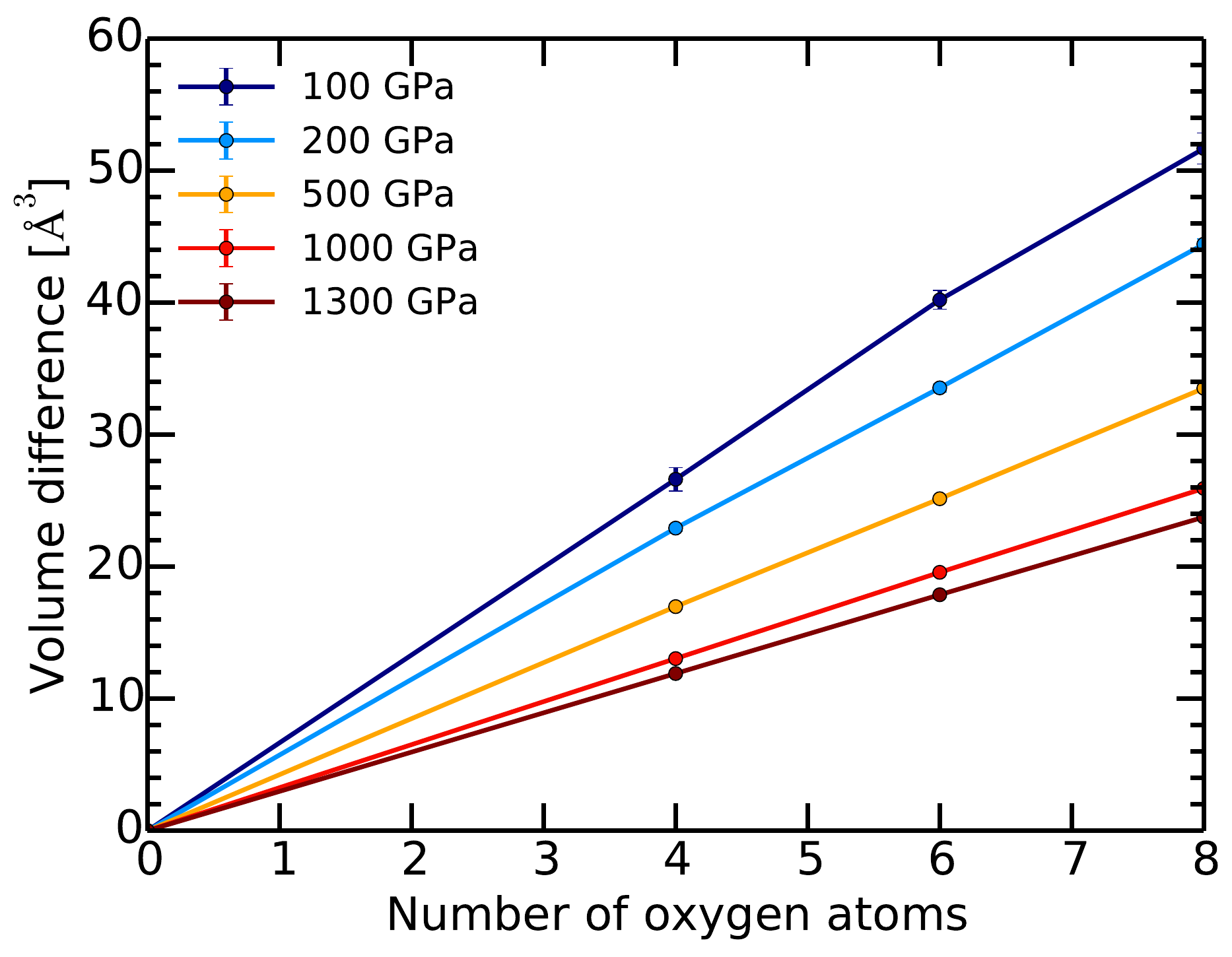}
\caption{\label{fig:volumespread} Volume difference between H-He-O and H-He 
mixtures as a function of the number of oxygen atoms for different pressures 
ranging from 100 to 1300~GPa and for a temperature of 10000~K.}
\end{figure}

In \figref{fig:volumetot}, we plotted the volume-pressure relationship for different 
H-He-O mixtures at 10000~K. For a given pressure, adding oxygen to the mixture 
results in an increase in the total volume. This effect is magnified in 
\figref{fig:volumespread} where we plotted the volume difference between the ternary 
and the binary mixtures as a function of the number of inserted oxygen atoms. For 
these conditions, the volume difference is an almost perfectly linear function of 
the number of oxygen atoms. This means that we can define an effective volume per 
oxygen atom in the H-He mixture, at given pressure $P$ and temperature $T$, by:
\begin{equation}
 v_\textrm{\tiny O,eff} = \frac{1}{N_\textrm{\tiny O}}\left[V(N_\textrm{\tiny 
H},N_\textrm{\tiny He},N_\textrm{\tiny O})-V(N_\textrm{\tiny H},N_\textrm{\tiny 
He})\right],
\end{equation}
where $V$ is the volume associated to a mixture of $N_\textrm{\tiny H}$ hydrogen, 
$N_\textrm{\tiny He}$ helium and $N_\textrm{\tiny O}$ oxygen atoms. Reciprocally, 
we can determine the volume for arbitrary but small concentration in oxygen by 
using an isothermal-isobaric additive mixing rule and the aforementioned effective 
volume of oxygen.

\begin{table}[b!]
\centering
\caption{Fit parameters of the effective volume as defined in 
\eqref{eq:vol}. The units of the parameters were chosen so that the temperature is in 
K, the pressure in GPa and the volume in \AA$^3$ per species.}
\label{table:fitvolparam}
{
\begin{tabular}{ccccc}
\hline\hline
 \multirow{2}{*}{Species} & $a$ & $b$ & $\alpha$ & $\beta$ \\
 & ($\times 10^{-3}$) &($\times 10$) & ($\times 10^{-6}$) & ($\times 10^{-1}$) \\
\hline
C & 1.154 & 1.899& -4.839& -2.890 \\
N & 1.000 & 2.093& -3.459& -2.990\\
O & 1.040 & 2.247& -4.119 & -2.916\\
Si & 2.901 & 3.219& -6.792& -2.915\\
Fe & 1.911 & 1.407& -8.600& -1.754\\
MgO & 4.277 & 6.302& -5.432& -3.185\\
SiO$_2$ & 4.180 & 8.119 & -4.411& -2.971\\

\hline\hline
\end{tabular}}
\end{table}

\begin{figure}[!t]
\centering
\includegraphics[width=0.98\columnwidth]{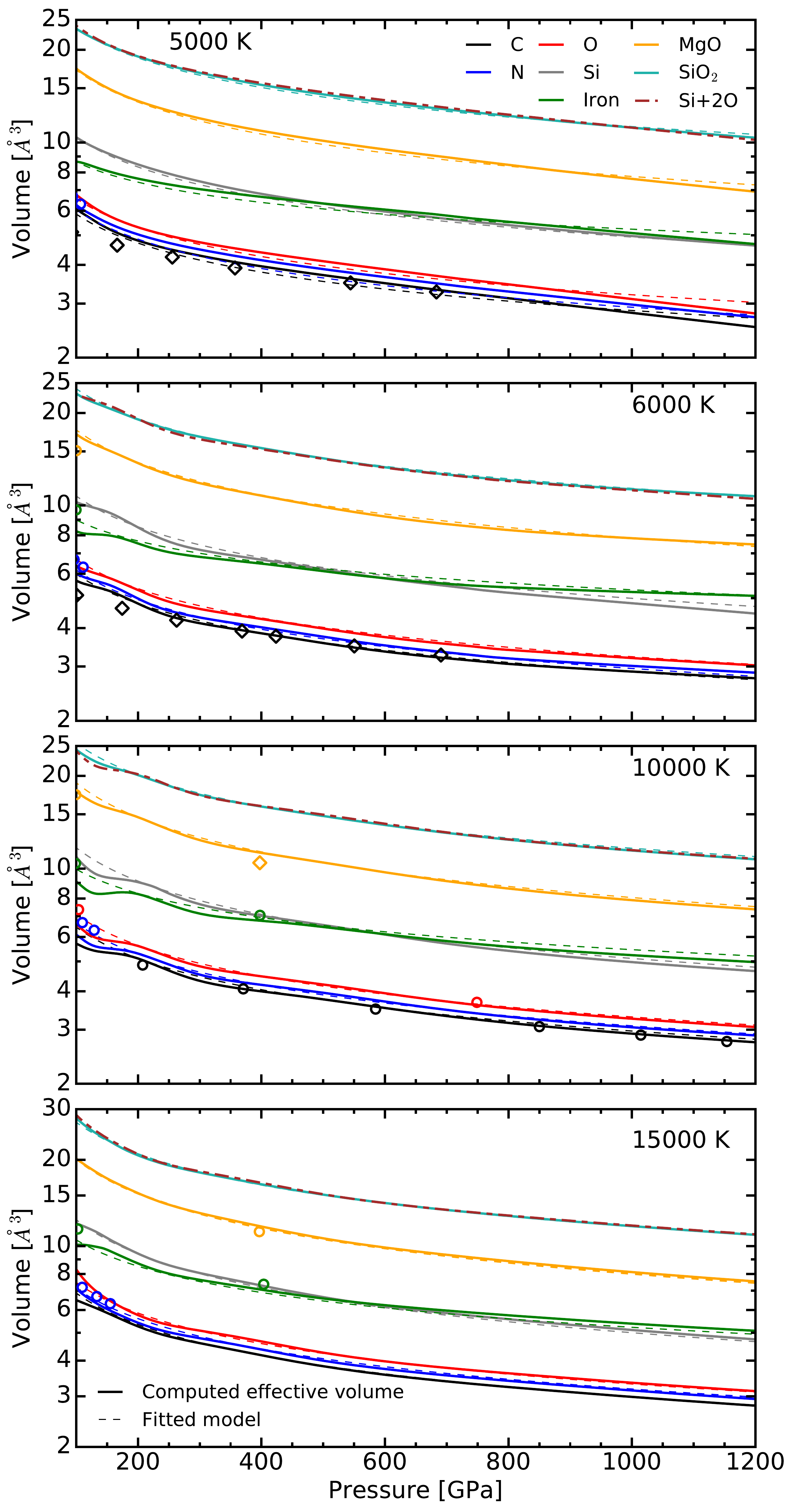}
\caption{\label{fig:volumeeff_high} Effective volume for each species in a H-He 
mixture as a function of the pressure for temperatures from 5000 to 15000~K. The 
full curves are the direct results from the volume difference analysis. The light green 
dash-dotted curve is the sum of the volumes of one Si and two O to be compared 
with the volume of one SiO$_2$. The dashed curves represent the fit based on 
\eqref{eq:vol} and the parameters given in \tab{table:fitvolparam}. The squares 
(resp. diamonds) represent the volume per species in a pure liquid (resp. solid) 
phase \citep{benedict_2014,driver_2016,driver_2015,wahl_2015}.}
\end{figure}

For pressures above 100~GPa and temperatures higher than 5000~K, we observed a 
linear behavior of the volume difference for each species we considered. In 
\figref{fig:volumeeff_high} we plotted the effective volume of each species that can 
be very well fitted as a function of both the temperature and the pressure with 
the following simple expression:
\begin{equation}
\label{eq:vol}
 v_\textrm{\tiny eff}(P,T) = (a\,T+b)\; P^{\alpha\,T+\beta},
\end{equation}
where $a$, $b$, $\alpha$ and $\beta$ are fit parameters given in 
\tab{table:fitvolparam}. To perform this fit we relied on results ranging from 100 
to 1500~GPa, from 5000 to 15000~K and for concentrations in heavy elements lower 
than 5\% in number. We first fitted the effective volume as a power-law of 
the pressure, along different isotherms, and using a weighted least-square fitting 
procedure. We then fitted the two coefficients of the power-law as an affine 
function of the temperature -- of the form $a\,T+b$. This provided a robust fit of 
the effective volume. This fit is an important result of this article because it 
describes the properties of heavy species under pressure-temperature conditions 
where H-He mixtures are metallic, which makes up for the major part of giant 
planet interiors.

We compared the effective volume with the volume of the pure species as available 
in the literature. We used data from \textit{ab initio} simulations for carbon 
\citep{benedict_2014}, nitrogen \citep{driver_2016}, oxygen \citep{driver_2015}, 
iron and magnesium oxide \citep{wahl_2015}. \figref{fig:volumeeff_high} shows a 
fairly good agreement between the volume of the pure species and their effective 
volume in H-He mixtures for pressure higher than about 200~GPa. This is not an 
obvious result as the interactions are mostly between the heavy species and 
hydrogen or helium in the ternary systems while they are only between the heavy 
elements themselves in the pure systems. On the other hand, below 200~GPa, some 
deviations have to be noted.   

We also compared the effective volume of an SiO$_2$ unit with the sum of the 
effective volumes of one silicon and two oxygen atoms in H-He mixtures. As shown 
on \figref{fig:volumeeff_high}, there is a very good agreement between the two 
estimates, which suggests that, under these conditions, the system is most likely 
dissociated. We infer that a similar behavior is to be expected to any 
multi-component system that would dissociates with increasing temperature and 
pressure.

\begin{figure}[!ht]
\centering
\includegraphics[width=0.98\columnwidth]{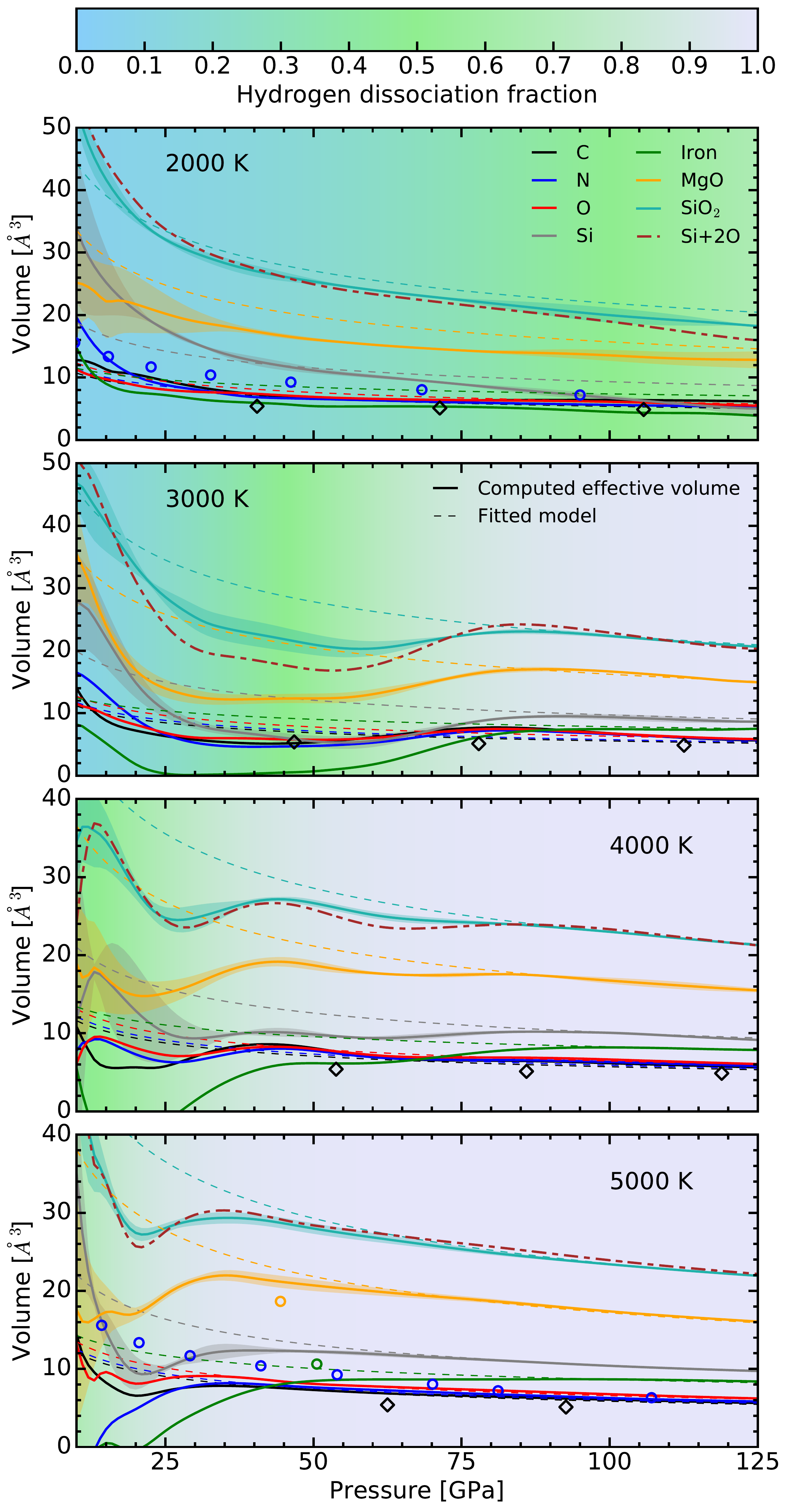}
\caption{\label{fig:volumeeff_low} Effective volume for each species in a H-He 
mixture as a function of the pressure for temperatures from 2000 to 5000~K. The 
legend is similar to \figref{fig:volumeeff_high}. The colored background represents 
the dissociation fraction of hydrogen in the H-He mixture (See \figref{fig:pie1} for 
the actual location of the dissociation in Jupiter). The shaded regions around the 
Si, MgO and SiO$_2$ curves show the estimated uncertainty.}
\end{figure}

We explored lower pressure-temperature range as well, from 10 to 150~GPa and 2000 
to 5000~K. In some cases for these conditions, the relationship between the volume 
difference and the number of the entities deviates from a perfect linear 
relationship. We think that these deviations may come from the finite duration of 
our simulations which may prevent to reach a perfect chemical equilibrium. We 
still determined an effective volume per species through a linear fit and the 
results are presented in \figref{fig:volumeeff_low}. To estimate the uncertainty on 
the effective volume, we combined the intrinsic statistical uncertainty of the 
effective volume with an estimate of the misfit. For the latter, we computed the 
effective volumes, $v_{\tiny\textrm{eff},i}=\Delta V(N_{\tiny Z,i})/N_{\tiny 
Z,i}$, from our simulations with different numbers of heavy species, $N_{\tiny 
Z,i}$. We then calculated their standard deviation from the linear fit value to 
characterize the misfit. While the statistical uncertainty is not very sensitive 
to the temperature or the pressure, the combined uncertainty of the effective 
volumes increases if the temperature decreases, which can be seen in 
\figref{fig:volumeeff_low}.

\begin{figure}[!ht]
\centering
\includegraphics[width=\columnwidth]{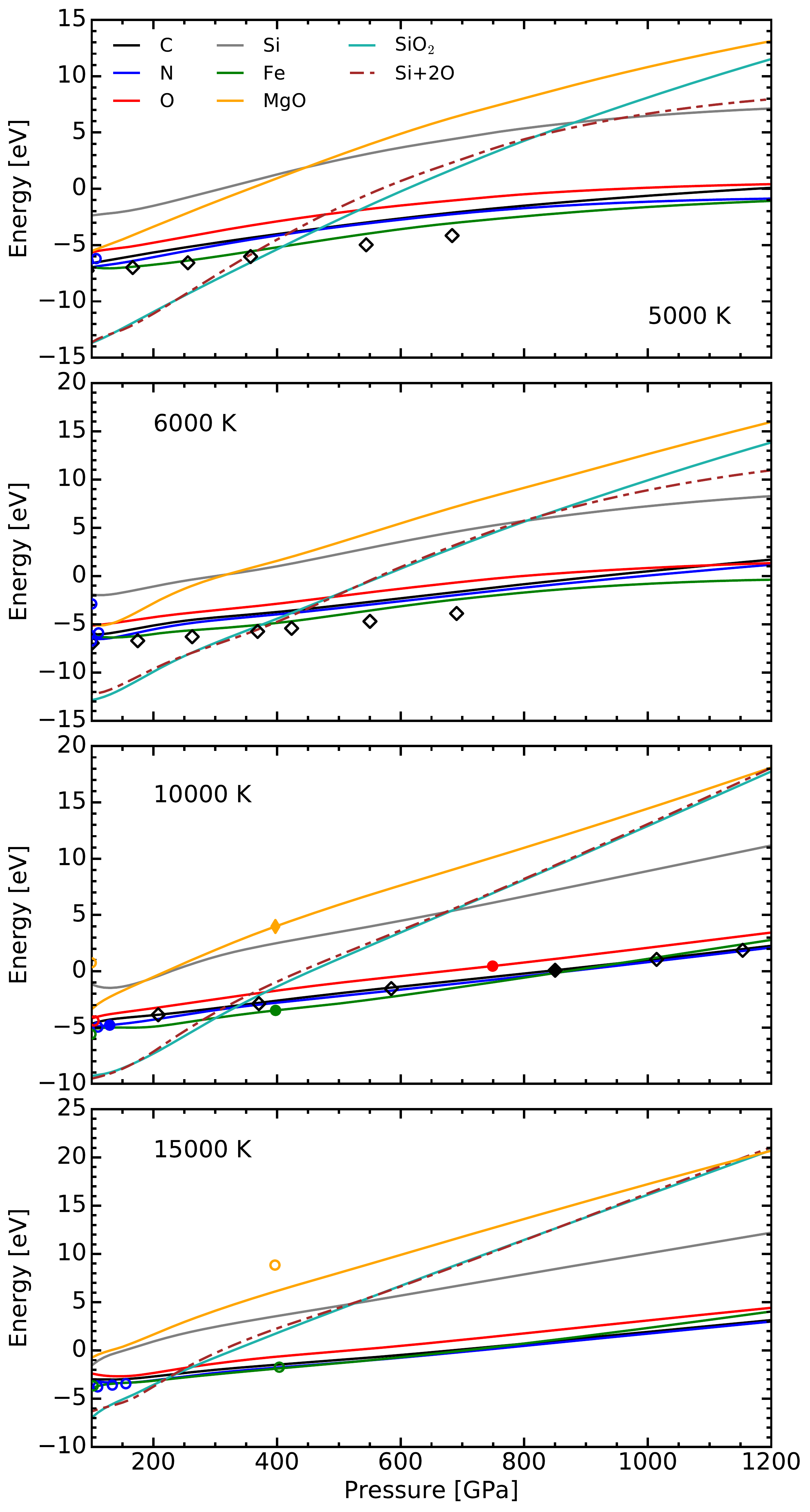}
\caption{\label{fig:energyeff_high} Effective energy for each species in a H-He 
mixture as a function of the pressure for temperatures from 5000 to 15000~K. The 
legend is similar to \figref{fig:volumeeff_high}. At 10000~K, the colored 
symbols represent the data for which we forced the energy to match our data to 
compensate for any difference in the energy reference.}
\end{figure}
\begin{figure}[!ht]
\centering
\includegraphics[width=\columnwidth]{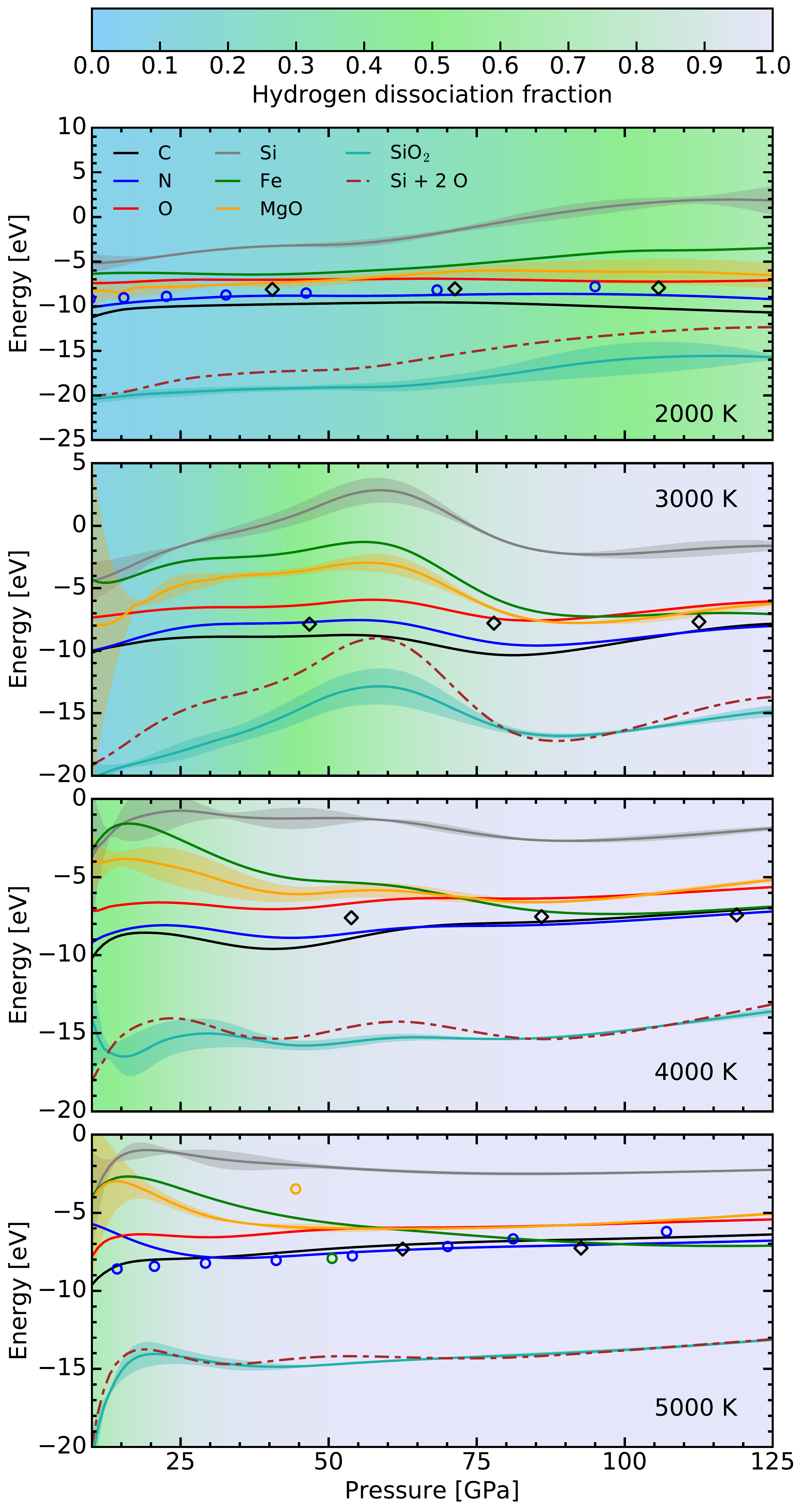}
\caption{\label{fig:energyeff_low} Effective energy for each species in a H-He 
mixture as a function of the pressure for temperatures from 2000 to 5000~K. The 
legend is similar to \figref{fig:volumeeff_low}.}
\end{figure}

Unlike for the high temperature and pressure conditions, the effective volume of 
each species does not evolve monotonically as a function of the pressure in the 
range of 10 to 150~GPa and 2000 to 5000~K, and an important variability can be 
observed. It is known that in this regime, hydrogen undergoes a dissociation and a 
metallization \citep{caillabet_2011,vorberger_2007}. By identifying the nearest 
neighbors over time as in \citet{soubiran_2015}, we determined an estimate of the 
dissociation fraction of hydrogen in the H-He mixtures. In 
\figref{fig:volumeeff_low}, we see a clear correlation between the drastic changes of 
the effective volume of the heavy species and the dissociation of hydrogen. This 
result is not surprising because the interactions between an H$_2$ molecule and a 
heavy atom are quite different from the interactions between an hydrogen ion and 
the same heavy element.

More specifically, the effective volume increases by nearly 25\% for C, N, O 
between 60 to 80~GPa at 3000~K, which could also be linked to chemical reactions 
with the surrounding hydrogen. At low temperature and pressure, for instance, 
carbon tends to form CH$_x$ molecules with $x$ ranging from 0 to 4 
\citep{sherman_2012} and oxygen associates to hydrogen to form hydroxide or water 
molecules \citep{soubiran_2014}. This chemistry can also explain the strong 
deviations of the effective volume of these species from the volume of the pure 
species. It is also interesting to note that iron has a negative effective volume 
at the lowest pressure conditions. We attribute this behavior to a complex 
chemistry with hydrogen. In the case of SiO$_2$ we also see a variability in the 
effective volume but more importantly, we see that there are some deviations from 
the sum of the effective volumes of one Si and two O atoms. This means that the 
system is not dissociated which modifies the interactions with the H-He mixtures. 

The high variability of the effective volume makes it impossible to give a simple 
fitting formula but we make our results on the effective properties available in 
the supplementary material attached to this article. Nevertheless, 
\figref{fig:volumeeff_low} shows that the formula from \eqref{eq:vol} with the 
parameters from Tab. \ref{table:fitvolparam}, fitted on the results above 5000~K 
and 100~GPa, actually well reproduces the lower-temperature behavior in the 
dissociated phase. Namely, at 5000~K, the fit gives reasonable values starting at 
50~GPa, and at 4000 and 3000~K, it gives accurate results from 80 to 150~GPa.

\begin{figure}[!t]
\centering
\includegraphics[width=\columnwidth]{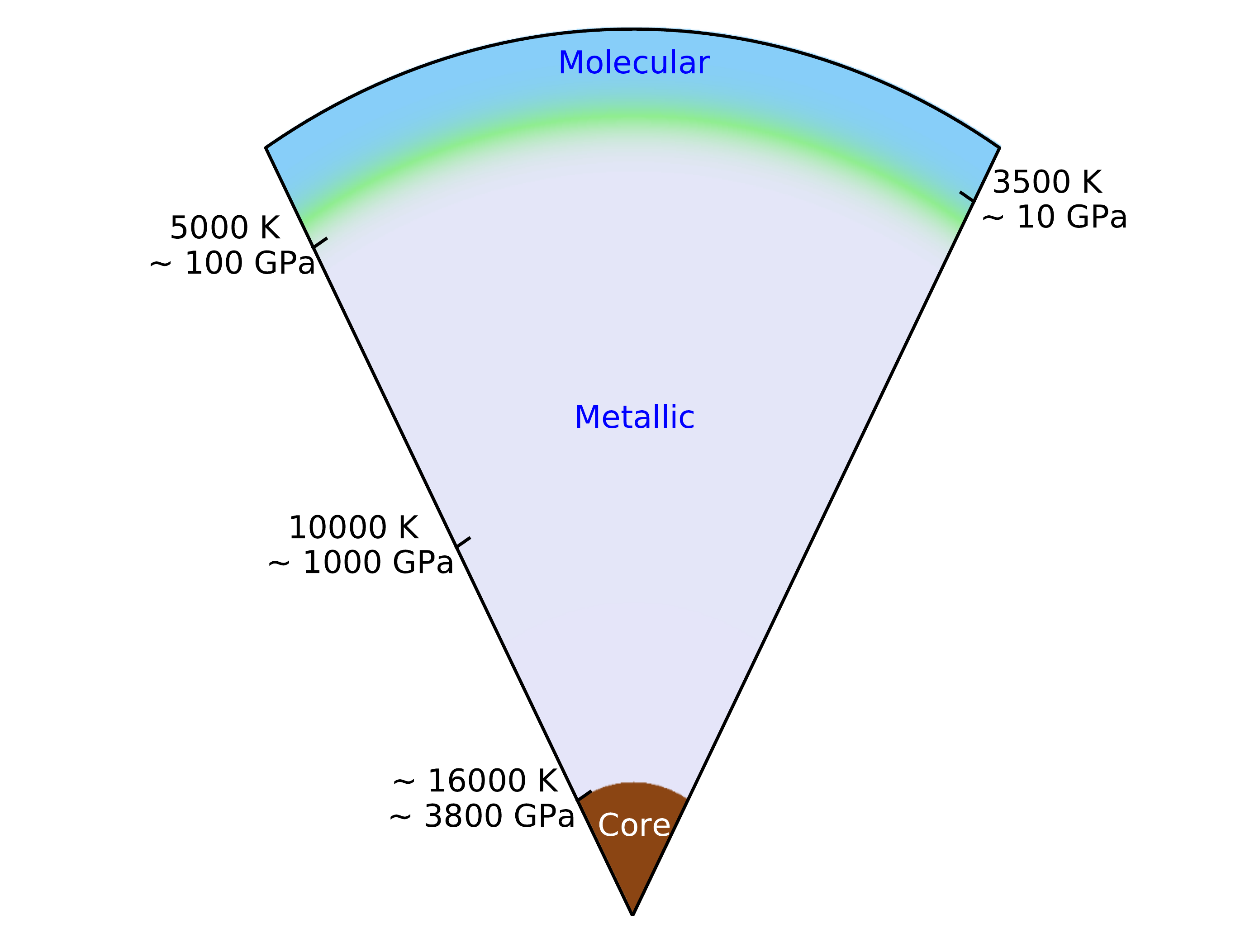}
\caption{\label{fig:pie1} Schematic view of the interior of Jupiter with the 
colors indicating the dissociation of hydrogen as in Figs. \ref{fig:volumeeff_low} 
and \ref{fig:energyeff_low}. We also indicated the order of magnitude of the 
thermodynamic conditions at different depths.}
\end{figure}

\subsection{Effective internal energy}
We investigated the effect of the inclusion of heavy elements in H-He mixtures on 
the internal energy of the systems. Since the energy is an extensive thermodynamic 
function like the volume, we followed the same procedure as in the previous 
section. The energy difference between the ternary and the binary mixtures 
exhibits a fairly linear behavior as a function of the number of heavy element 
entities (graph not shown). By comparing the energy of the ternary mixture of 
$N_\textrm{\tiny H}$ hydrogen atoms, $N_\textrm{\tiny He}$ helium atoms and 
$N_\textrm{\tiny Z}$ Z entities (atoms or molecules) with the energy of the binary 
H-He mixture at the same pressure $P$ and temperature $T$, we were able to 
determine an effective energy per species Z:
\begin{equation}
 e_\textrm{\tiny Z,eff} = \frac{1}{N_\textrm{\tiny Z}}\left[E(N_\textrm{\tiny 
H},N_\textrm{\tiny He},N_\textrm{\tiny Z})-E(N_\textrm{\tiny H},N_\textrm{\tiny 
He})\right].
\end{equation} 

Our results for the effective energies are plotted in \figref{fig:energyeff_high} and 
\ref{fig:energyeff_low}. In the dissociated regime, the effective energy exhibits 
a fairly smooth evolution with low deviations from the linear behavior. We 
compared the effective energy with the internal energy of the pure species. In 
order to correct for any shift in the origin of the energy between the different 
data sources, we artificially made them to coincide at 10000~K, temperature at 
which we had the most data and where we expect every species to be dissociated. 
The full symbols in \figref{fig:energyeff_high} indicate the specific data we forced 
to match the effective energies we computed. This results in an artificially good 
match on the 10000~K isotherm. However, if we look at the other temperatures, we 
observe some strong deviations between the effective energy and the pure species 
energy, emphasizing the importance of the inter-species interactions. Below 
5000~K, the effective energy exhibits similar features to the effective volume 
with drastic variations as a function of the pressure when hydrogen undergoes its 
dissociation. 

Overall, we observe that the energy of the ternary mixtures can be approximated by 
an isothermal-isobaric additive mixing rule which is very helpful for evolution 
models of planets. However, the effect of the dilution of the heavy species into 
the H-He cannot be neglected, since the effective energy is substantially 
different from the internal energy of the pure species.

\subsection{Effective entropy\label{sec:entropy}}
In order to determine the influence of the heavy elements on the 
temperature-pressure profile of the giant planets, one needs to compute the 
entropy of the ternary mixtures. Hence, we computed the entropy of ternary 
mixtures using thermodynamic integrations. As the computation cost of such 
calculations is very high, we only performed it for oxygen in H-He mixtures and 
for a subset of temperatures. We shall see in Section \ref{sec:isentrope} that it 
is already enough to determine the effect on the density profile. 

We followed the same procedure as for the volume and the energy: we computed the 
entropy of the mixture for different concentrations and compared the results with 
the binary H-He mixture. However, the entropy encompasses not only the entropy of 
each species but also a mixing entropy. We therefore defined the effective 
entropy 
based on the total entropy of the ternary mixtures of $N_\textrm{\tiny H}$ 
hydrogen atoms, $N_\textrm{\tiny He}$ helium atoms and $N_\textrm{\tiny O}$ 
oxygen 
atoms and the entropy of the H-He binary mixture, at constant pressure $P$ and 
temperature $T$, by:
\begin{eqnarray}
 s_\textrm{\tiny O,eff} &=& \frac{1}{N_\textrm{\tiny O}}[S(N_\textrm{\tiny 
H},N_\textrm{\tiny He},N_\textrm{\tiny O})-S(N_\textrm{\tiny H},N_\textrm{\tiny 
He}) \nonumber \\
 &-& \Delta S_\textrm{\tiny mix}(N_\textrm{\tiny H},N_\textrm{\tiny 
He},N_\textrm{\tiny O})], \label{eq:entropy}
\end{eqnarray} 
where the entropy of mixing is given by:
\begin{eqnarray}
 \Delta S_\textrm{\tiny mix}(N_\textrm{\tiny H},N_\textrm{\tiny 
He},N_\textrm{\tiny O})&=&k_\textrm{\tiny B}\,\ln(N_\textrm{\tiny 
H}+N_\textrm{\tiny He}+N_\textrm{\tiny O})!\nonumber\\
 &-&k_\textrm{\tiny B}\,\ln(N_\textrm{\tiny H}+N_\textrm{\tiny He})! \nonumber\\
 &-&k_\textrm{\tiny B}\,\ln N_\textrm{\tiny O}!.
\end{eqnarray}
We chose this formula with the explicit factorial term because the usual Sterling 
approximation is not appropriate for 
small numbers. 

\begin{figure}[!t]
\centering
\includegraphics[width=\columnwidth]{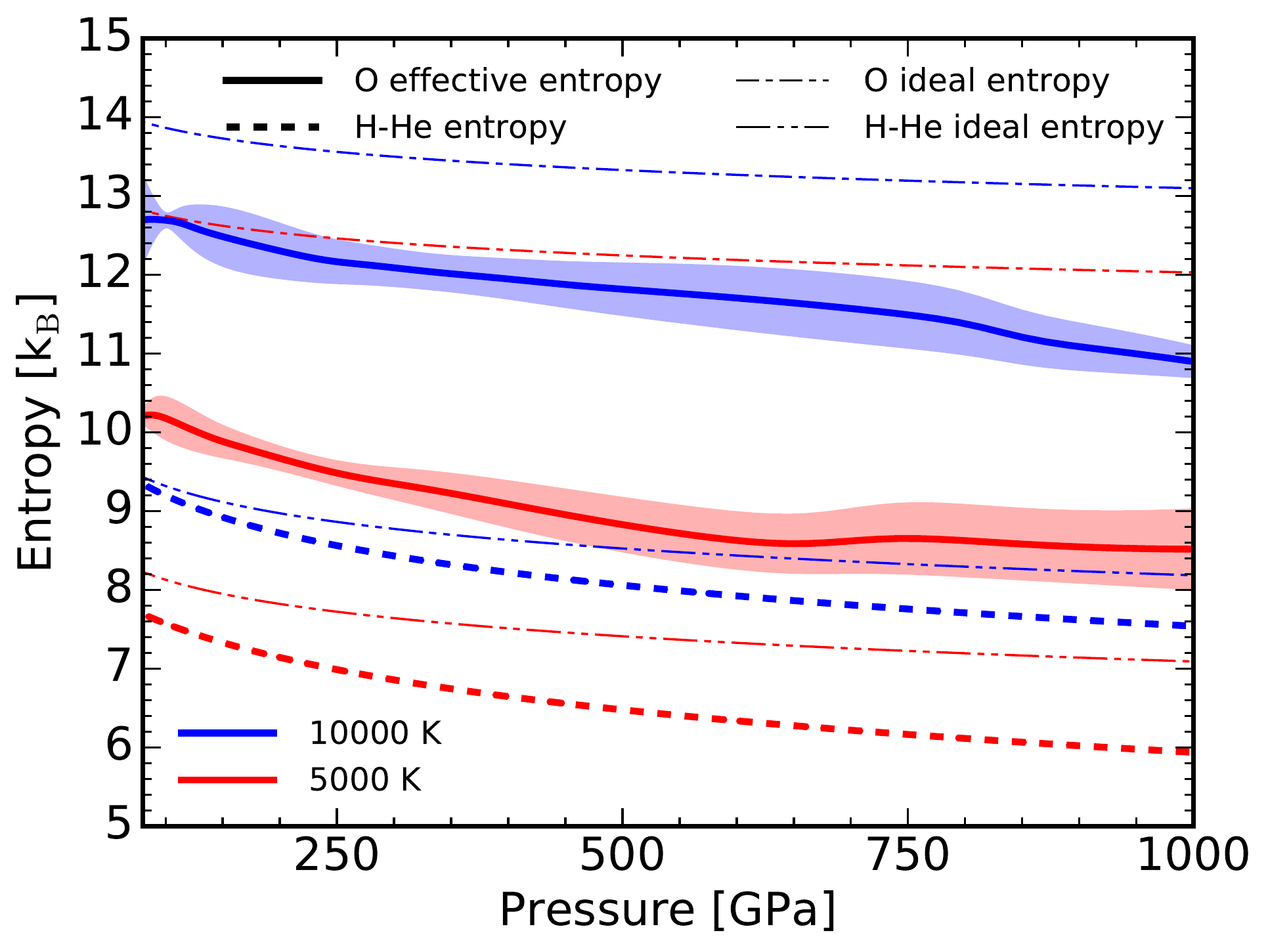}
\caption{\label{fig:entropy} Effective entropy per oxygen atom in a ternary H-He-O 
mixtures as a function of the pressure, for 5000 and 10000~K. The shaded region 
represents our estimate of the uncertainty. We also plotted in dashed 
lines the entropy per nucleus for the binary H-He mixture under the same 
conditions. For comparison, the thin lines show the ideal entropy of pure oxygen 
and of the H-He mixture under equivalent density and temperature conditions.}
\end{figure}

We plotted the effective entropy of oxygen in \figref{fig:entropy}. The 
uncertainty is slightly higher than on the energy or the volume because the 
entropy calculation requires more computation steps but it remains very reasonable 
overall, as the maximum uncertainty does not exceed 6~\%. We also plotted the 
entropy per nucleus of the binary H-He mixtures under the same conditions. For 
comparison we computed the ideal entropy of the H-He mixture and of oxygen using 
the Sackur-Tetrode formula \citep{Reichlbook}:
\begin{equation}
 S_\textrm{\tiny id}=k_\textrm{\tiny B}\,\sum_\alpha 
x_\alpha\left(\ln\left[v_\alpha\left(\frac{ m_\alpha k_\textrm{\tiny B} T }{2\pi 
\hbar^2}\right)^{3/2}\right]+\frac{5}{2}\right),
\end{equation}
where $x_\alpha$, $v_\alpha$ and $m_\alpha$ are the concentration ratio, 
volume per particle and mass per particle of the particles of type $\alpha$ 
respectively. We used the $P$-$V$ relationship of the H-He mixture to derive the 
ideal entropy of H-He as a function of the pressure. For oxygen, we based our 
calculation on the effective volume as derived in section \ref{sec:effvol}.

The effective entropy of oxygen is higher than the entropy of H-He, which 
is simply a mass effect, which is present in the ideal entropy. Both the effective 
entropy of oxygen and entropy of H-He are smoothly decreasing as pressure 
increases because the volume per nucleus decreases. The non-ideal effects -- the 
difference between actual and ideal entropy -- increase with pressure because the 
interactions introduce some local order. The entropy and effective entropy 
increase as temperature increases consistently with the ideal entropy although the 
increase is lower for the ideal entropy. We also note that as the temperature 
increases, the non-ideal effects decreases, which is consistent with a diminution 
of the interactions effects. We observe that the non-ideal effects appear more 
pronounced on the oxygen than on the H-He mixture. But we have to stress that this 
is here only an effective entropy of a single oxygen atom in a H-He fluid and that 
the entropy of pure oxygen under similar conditions may be quite different. The 
interactions with hydrogen and helium may also influence the local ordering around 
the oxygen atoms modifying the entropy. Last, we want to stress that the 
variations in entropy or effective entropy as a function of the pressure and the 
temperature are quite similar for H-He and for oxygen, which is important when 
determining the internal profile of a giant planet.

\section{Discussion}\label{sec:isentrope}
The analysis of the \textit{ab initio} simulations of several ternary mixtures 
showed that, in the diluted regime, the addition of heavy elements to an H-He 
mixtures can be very well described by an ideal isothermal-isobaric additive 
mixing rule as long as we employ the effective properties of the heavy materials 
as presented in the previous section. For the higher pressures in the dissociated 
regime, the effective volume coincides with the volume of the pure systems but 
deviations are observed for lower pressures emphasizing the need for these 
effective properties. 

In the diluted limit, we can also expect that we can approximate the properties of 
a multi-component system by simply adding the effective volumes or energies of 
each component separately, because the cross-interactions between different heavy 
elements should be negligible compared to the interactions with hydrogen and 
helium. This is further supported by the good agreement between the SiO$_2$ 
effective properties and those of silicon and oxygen taken separately. This 
approximation is however only accurate for the dissociated regime, which actually 
represent a significant mass fraction of the interiors of Jupiter and Saturn. All 
the previous calculations were performed for a fixed hydrogen to helium ratio, but 
the deviations from the pure species properties are small enough to let us believe 
that the deviations on the effective properties, induced by a reasonable change in 
helium concentration, should be mostly negligible.

\begin{figure}[!t]
\centering
\includegraphics[width=\columnwidth]{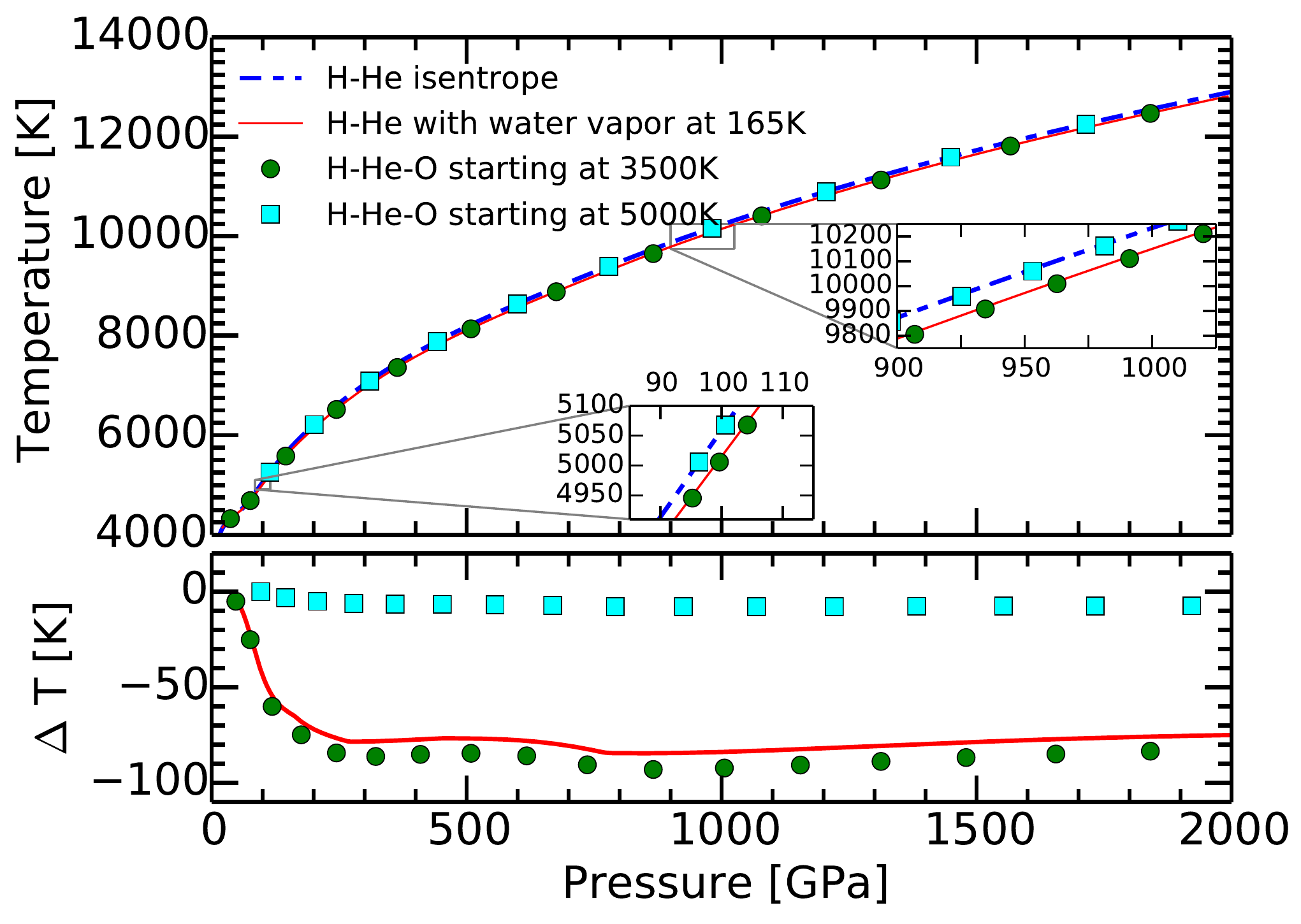}
\caption{\label{fig:isent} Pressure-temperature profile in the H-He envelope of a 
gaseous planet with $T=165$~K at 1~bar with a 2\% mass fraction of oxygen. The 
full and dashed lines are for a single layer of H-He-O with or without entropy 
correction due to the presence of oxygen. The discrete symbols are for a two-layer 
picture, where the 2\% oxygen enriched layer starts at 3500 or 5000~K. A summary 
of the properties of these profiles can be found in \tab{table:isentrope}. The 
bottom graph shows the temperature difference between the different profiles with 
the original H-He profile.}
\end{figure}

Using the results of our calculations, we studied the consequences of adding heavy 
elements in the hydrogen and helium rich envelope of giant planets. As a toy 
model, we considered as a starting point an homogeneous fully convective H-He 
envelope with a composition of H:He=220:18 as employed in our simulations. We 
picked the temperature at 1 bar to be $T=165$~K, which is close to the Jupiter 
measured value of 166.1~K \citep{seiff_1998}, for instance. With this condition 
and using \citet{militzer_2013} and \citet{saumon_1995} EOSs, we were able to 
determine the pressure-temperature profile, plotted in \figref{fig:isent}. We also 
computed the density along the isentrope $\rho\indice{H-He}(P)$.  

With the effective volumes in \tab{table:fitvolparam}, we computed the excess 
density induced by the addition of a $Z=0.02$ mass fraction of different heavy 
elements homogeneously throughout the H-He envelope. The pressure-temperature 
profile was kept constant as in \citet{hubbard_2016}, and we only perturbed the 
density. 

\begin{figure}[!t]
\centering
\includegraphics[width=\columnwidth]{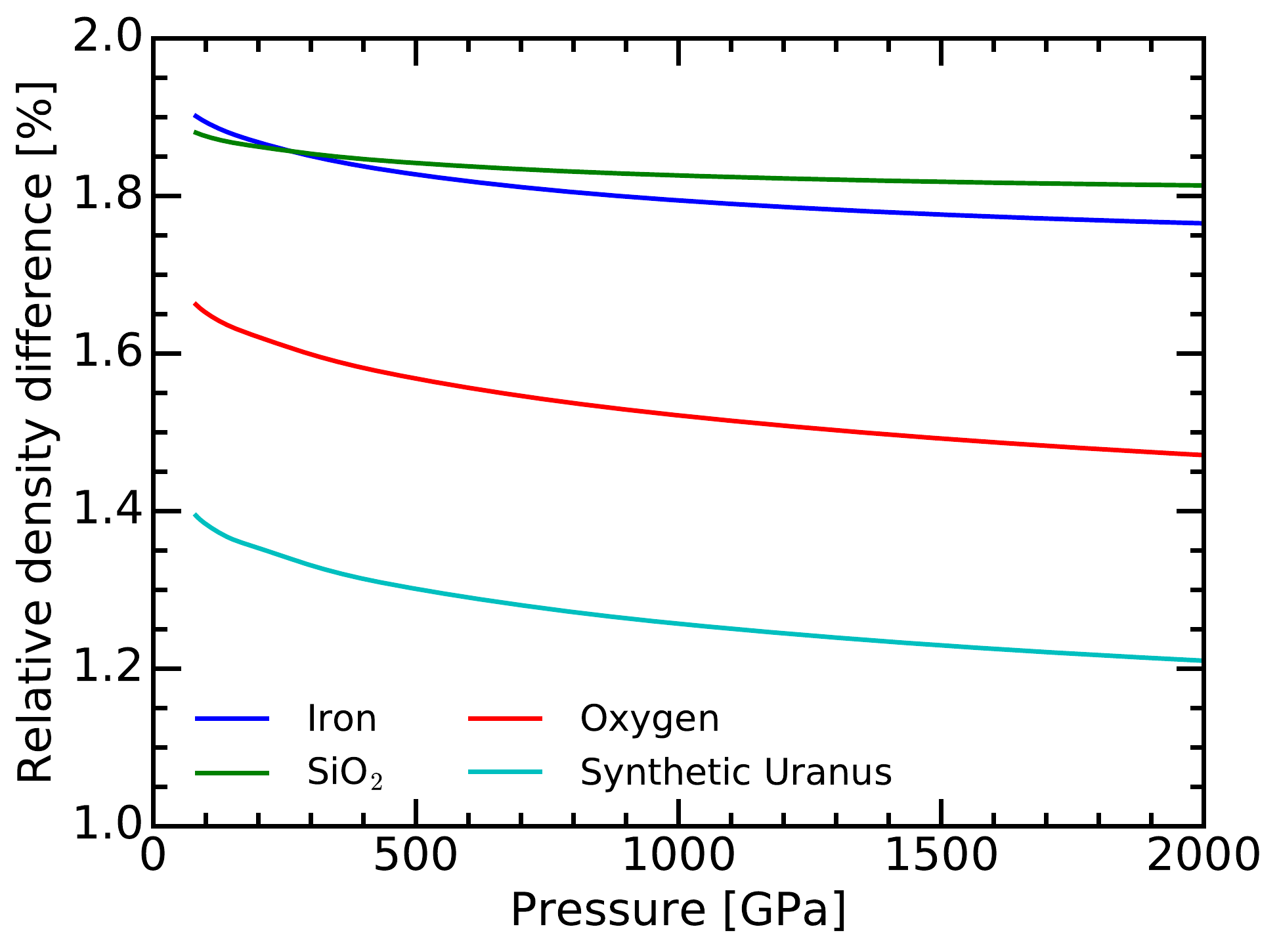}
\caption{\label{fig:barotrope} Relative density difference, along the H-He 
isentrope of a giant planet with $T=165$~K at 1~bar, between a simple H-He mixture 
and a multi-component mixture including a 2~\% mass fraction of heavy elements. 
Synthetic Uranus refers to a mixture of ice derivatives as presented in 
\citet{nellis_1997,hubbard_2016}. For this species, we chose a composition of 
H:O:C:N=87:25:13:4.}
\end{figure}

In \figref{fig:barotrope}, we displayed the relative density difference between the 
enriched envelope and the original H-He envelope 
$[\rho\indice{H-He-Z}(P)-\rho\indice{H-He}(P)]/\rho\indice{H-He}(P)$. The curves 
are all for the same mass fraction, $Z=0.02$, but for different materials. They 
illustrate by how much the multi-component mixture deviates from the H-He mixture. 
In the most extreme case, when heavy atoms have a negligible volume, a 2\% mass 
fraction leads to a 2\% density change. As expected, the densest species -- iron 
and silicon dioxide -- introduce the largest density change, close to a 2\% 
density increase for a 2\% mass fraction. On the other hand, the inclusion of 
Synthetic Uranus (SU) induces a 1\% density increase only. SU is a proxy that we 
used to mimic a mixture of ice derivatives based on oxygen, nitrogen, carbon and 
hydrogen in a ratio of H:O:C:N=87:25:13:4. It was first introduced as a proxy for 
Uranus interior in high pressure laboratory experiments \citep{nellis_1997}, and 
it was also used in the recent Jupiter model by \citet{hubbard_2016} to enrich the 
envelope. The relatively low density of SU is due to the presence of a high amount 
of hydrogen that we chose to include in the computation of the $Z$ fraction. For 
comparison, oxygen has a slightly higher density with a density increase of 
roughly 1.5\%.

We showed that the addition of heavy elements in the H-He mixture slightly 
influenced the density of the mixture. It is thus natural to infer that they may 
have an influence on the entropy as well, and thus, on the pressure-temperature 
profile of the envelope. Yet, a change in the $P$-$T$ relationship in the planet 
would also change its density profile and it has to be accounted for. Besides the 
case of the unperturbed H-He isentrope in \figref{fig:isent}, we explored three 
different scenarios for the enriched layer. The first scenario is a single layer 
of a H-He-O mixture with a $Z=0.02$ mass fraction in oxygen but this time taking 
the entropy change into account. We then explored two other scenarios with a 
two-layer picture where the outer layer is made of H-He solely and the inner layer 
is an enriched H-He mixture with 2\% of oxygen. The difference lies in the 
temperature at the interface between the two layers: in one case we considered a 
3500~K interface, in the other we picked 5000~K. The resulting properties for the 
different scenarios are summarized in \tab{table:isentrope}.

\begin{table}[b!]
\centering
\caption{Properties of the hypothetical isentropes in an H-He 
envelope enriched with 2\% oxygen in mass and with a 165~K temperature at 1~bar. 
The first two models are for a single layer without or with entropy correction. 
The last two models are for a two-layer picture with the enriched layer only for 
temperatures higher than 3500 or 5000~K. For each hypothesis, we give the entropy 
per nucleus $S$ as well as the pressures at 5000 and 10000~K. }
\label{table:isentrope}
{\scriptsize
\begin{tabular}{cccc}
\hline\hline
\multirow{2}{*}{Description} &   $S$ & $P_\textrm{\tiny 5000K}$ & $P_\textrm{\tiny 10000K}$ \\
 & (k$_\textrm{\tiny B}$/nucl.) & (GPa) & (GPa) \\
\hline
Unperturbed H-He & 7.598 & 95.6 & 927  \\
H-He-O mono-layer & 7.607& 98.6& 950\\
H-He-O starts at 3500~K & 7.605& 98.9& 953\\
H-He-O starts at 5000~K &  7.624& 95.6& 929\\
\hline\hline
\end{tabular}}
\end{table}

For the first scenario, we assumed that the oxygen reacted with the surrounding 
hydrogen at 165~K and 1~bar to form water molecules. We further assumed that water 
is in a vapor state as the concentration is low. We then could compute the entropy 
of the mixture at 165~K and 1~bar by:
\begin{eqnarray}
 s&=&x_{\textrm{\tiny H}_2} s_{\textrm{\tiny H}_2} + x_{\textrm{\tiny He}} 
s_{\textrm{\tiny He}} + x_{\textrm{\tiny H}_2\textrm{\tiny O}} s_{\textrm{\tiny 
H}_2\textrm{\tiny O}}\nonumber \\
 &-& k_\textrm{\tiny B} (x_{\textrm{\tiny H}_2} \ln x_{\textrm{\tiny H}_2} + 
x_{\textrm{\tiny He}} \ln x_{\textrm{\tiny He}} + x_{\textrm{\tiny 
H}_2\textrm{\tiny O}} \ln x_{\textrm{\tiny H}_2\textrm{\tiny O}}), 
\end{eqnarray}
where $x_\alpha$ (resp. $s_\alpha$) is the number fraction (resp. entropy per 
molecule) of molecules of type $\alpha$. We used \citet{saumon_1995} EOS to compute 
the entropy of H and He. For water, we used the entropy from the translational, 
rotational and vibrational degrees of freedom as in \citet{vidler_2000}.We 
obtained an entropy of 7.607~k$_\textrm{\tiny B}$/nucleus for the whole mixture.

\begin{figure}[!t]
\centering
\includegraphics[width=\columnwidth]{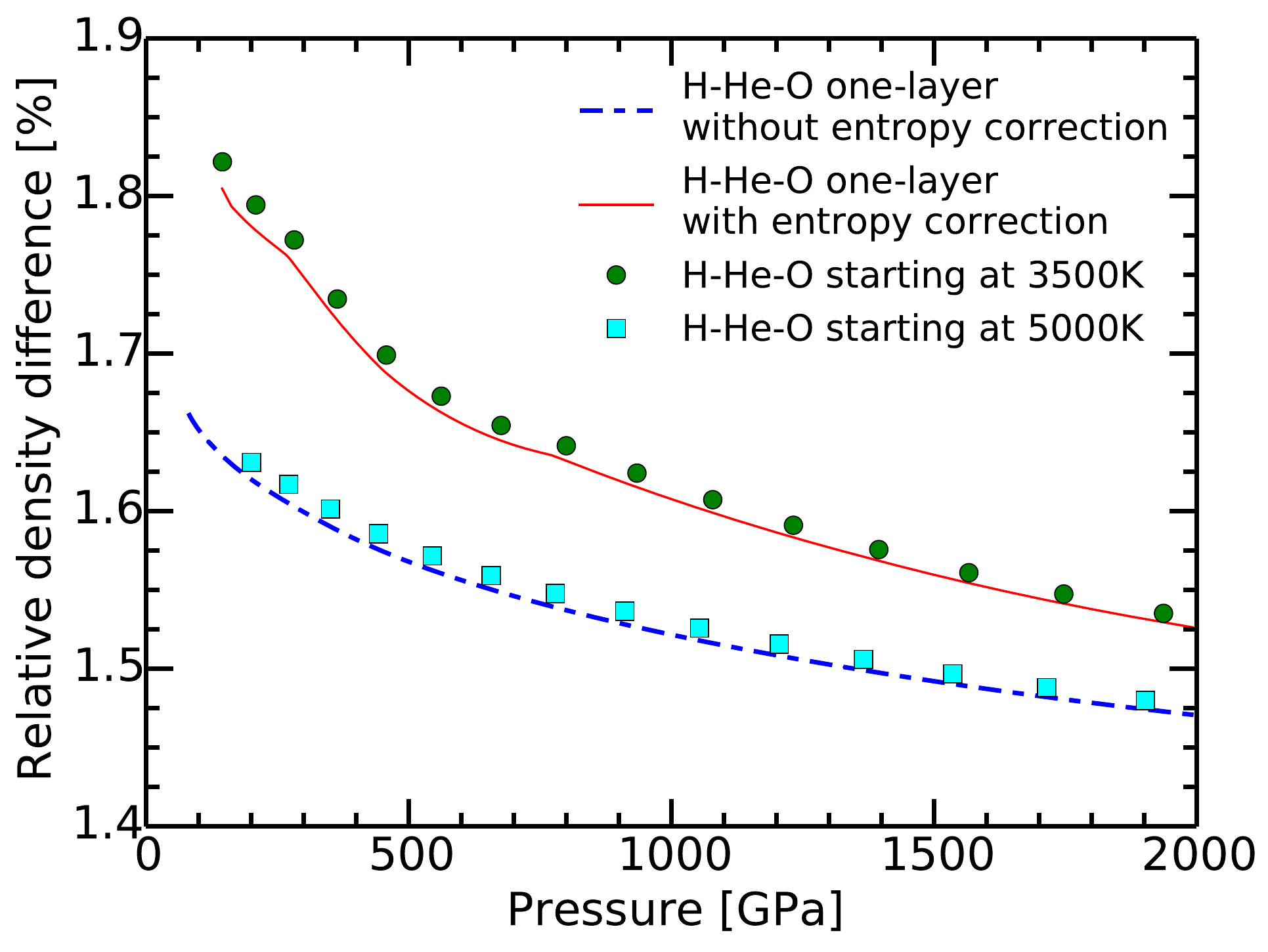}
\caption{\label{fig:isentdiff} Relative density difference between a simple H-He 
mixture and a multi-component mixture including a 2\% mass fraction of oxygen 
along the $P$-$T$ profile of the envelope of a giant planet with a 165~K temperature 
at 1 bar. The dashed blue (resp. solid red) curve shows the density increase for a 
single layer without (resp. with) the entropy correction due to the oxygen. The 
green circles (resp. cyan squares) are for a two-layer hypothesis with an outer 
H-He layer and an inner H-He-O layer starting at 3500~K (resp. 5000~K). The 
corresponding $P$-$T$ profiles are those of \figref{fig:isent}.}
\end{figure}

Following a constant entropy line, and using the effective entropy of oxygen 
computed in Section \ref{sec:entropy}, we were able to compute the$P$-$T$ 
relationship above 4000~K and 10~GPa, as represented in \figref{fig:isent}. The 
comparison with the unperturbed isentrope shows that the oxygen entropic effects 
result in an increase of the pressure by 3\% for a given temperature. 
Equivalently, it results in a temperature decrease of roughly 80~K at constant 
pressure. The effect on the density can be seen in \figref{fig:isentdiff}: compared 
to the original H-He profile, the density variation induced by the oxygen 
increases from 1.5\% without the entropy correction to 1.6\% when taking the 
entropy of oxygen into account. Since this is a rather marginal effect, it is hard 
to believe that this difference could be resolved by Juno measurements of the 
gravitational moments. We find it reasonable to use the original H-He isentrope 
for giant planet modeling \citep{hubbard_2016}.

The second and third scenarios are based on a two-layer picture with a pure H-He 
outer envelope and an inner envelope with a 2\% mass fraction of oxygen. The outer 
layer has the unperturbed $P$-$T$ profile and the enriched inner layer profile is 
determined by the condition at the interface: we computed the entropy of the 
ternary H-He-O mixture assuming that, at the interface, the pressure and the 
temperature were the same in both layers. The resulting profiles are represented 
by the symbols in Figs. \ref{fig:isent} and \ref{fig:isentdiff}.    If we let the 
inner layer start in the molecular phase, at 3500~K and 7.2~GPa, we retrieve to a 
very good accuracy the predictions of the single layer. It is mostly the 
dissociation that drives the slight shift in the isentrope and it occurs at 
temperatures higher than 3500~K. On the other hand, if we let the inner layer 
start at 5000~K and 95.6~GPa, in the dissociated phase, the predicted isentrope is 
almost exactly the same as the H-He isentrope. The temperature difference is lower 
than 10~K at a given pressure and the pressure is increased by only 0.2\% at 
10000~K (see \tab{table:isentrope}). The entropy itself is modified by the 
presence of oxygen, but this shift is nearly constant along the H-He isentrope in 
the dissociated phase, which explains the absence of impact on the predicted 
pressure-temperature profile and thus on the density. We expect this observation 
to be true for the other heavy element as well, especially in the diluted limit. 

The last scenario with the temperature at the interface at 5000~K is of interest 
for cold giant planet modeling. When the planet is cold enough, a phase separation 
of hydrogen and helium is indeed expected to occur, naturally differentiating the 
envelope in (at least) two layers. The innermost layer and helium enriched is 
entirely in the dissociated regime \citep{guillot_2005,hubbard_2016} and is also 
the one the most subject to heavy element enrichment because in direct contact 
with the eroding core. Yet, we showed on the example of oxygen that in this 
regime, the entropy of the heavy element play virtually no role on the 
pressure-temperature profile that can thus be determined by the H-He properties 
solely. This means that to model this innermost layer, the H-He EOS and the 
effective volumes of the heavy elements are sufficient to properly recover the 
density profile.

\section{Mass-Radius Relationship}

\begin{figure}[!t]
\centering
\includegraphics[width=\columnwidth]{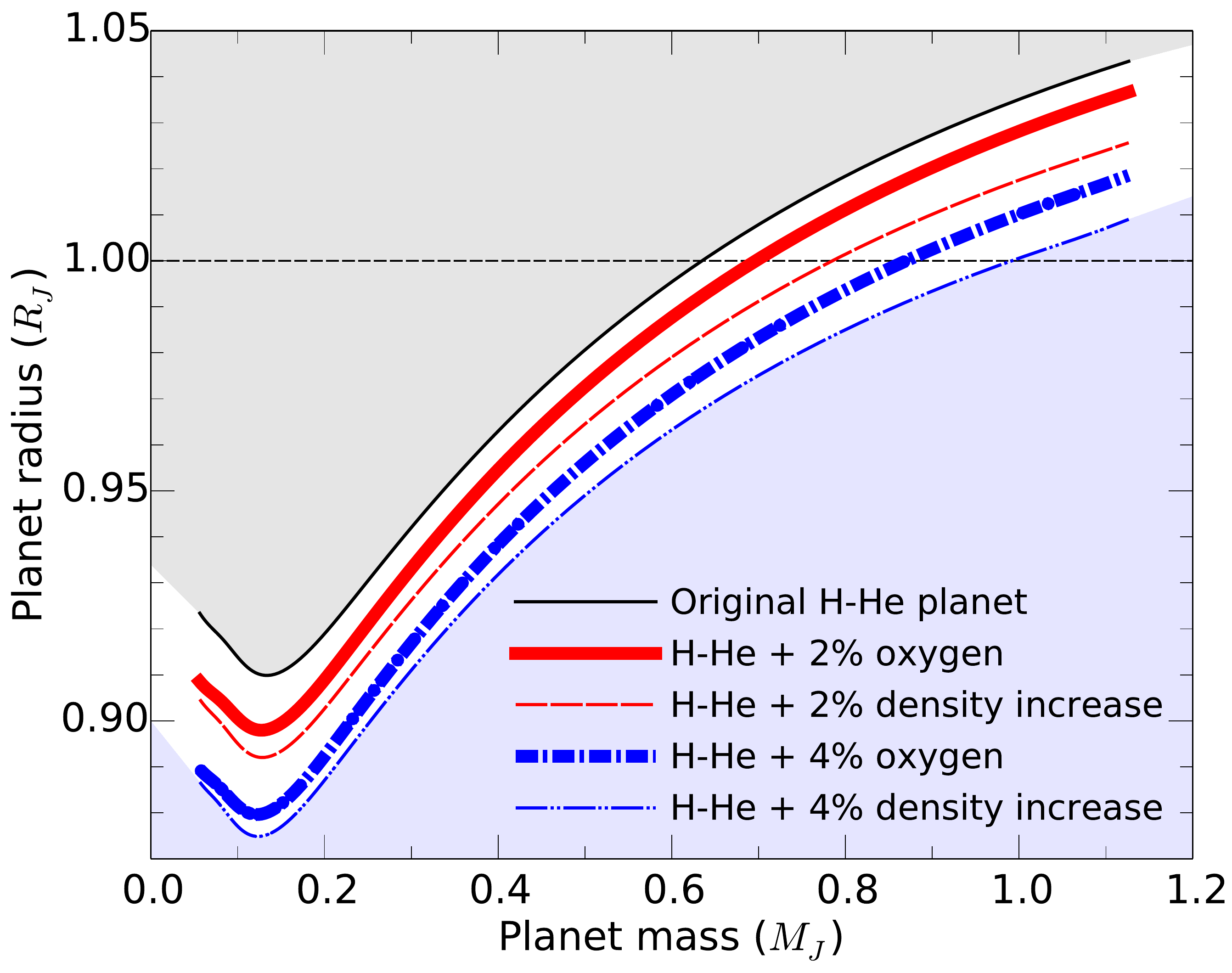}
\caption{Mass-radius relationship in Jupiter units for giant planets without a 
rocky core. The adiabatic interior $P$-$\rho$-$T$ profiles are derived from H-He 
mixture with an effective temperature of 166.1 K at 1 bar. The thin solid line 
shows the radii of the H-He planets without heavy elements. The thin dashed lines 
illustrate the effect that a 2\% or 4\% increase in density has on the planet 
radius.  The thick lines show the effect of introducing a 2\% and 4\% mass 
fraction of oxygen.  }
\label{fig:MR} 
\end{figure}

In this section, we briefly discuss the effects that heavy elements have on the 
mass-radius relationship of gas giant exoplanets. To simplify our analysis, we 
only consider planets without a rocky core and assume a homogeneous, convective 
interior. The adiabatic $P$-$T$ profile is derived from a H-He mixture starting 
from 1~bar and 166.1~K. For a given planet mass, the radius is derived by solving 
the ordinary differential equations of the hydrostatic 
equilibrium~\citep{seager_2007,wilson_2014}. \figref{fig:MR} shows the effect that 
the introduction of a 2\% and a 4\% mass fraction of oxygen has on the radius of a 
giant planet. We find that a Jupiter-mass planet respectively shrinks by 0.7\% and 
2.4\% in radius. According to \eqref{eq:vol}, the introduction of oxygen not only 
increases the mass but also the volume of the H-He mixture at given pressure and 
temperature. As a result, the density increases by less than 2\% or 4\%; 
respectively.  In order to disentangle both effects, we plot the mass-radius 
relationship for planets where we have increased the H-He density by 2\% and 4\% 
in \figref{fig:MR}. We find that a Jupiter-mass planet then shrinks by 1.7\% and 
3.3\% in radius; respectively.

It is interesting to note that the effect of heavy elements is much larger if one 
compares different planet masses for a given radius. Since giant planets have 
degenerate interiors, their radii start to shrink if a mass of approximately 3 
Jupiter masses is exceeded. Jupiter's radius is not too different from the maximum 
radius of approximately 1.2 $R\indice{J}$~\citep{seager_2007}. The slope of the 
mass-radius curves in \figref{fig:MR} is thus relatively small and a modest change in 
radius has a significant effect on the inferred planet mass. For a fixed radius of 
1 $R\indice{J}$, the mass of a core-less planet increases from 0.63 to 0.70 or 
0.86 $M\indice{J}$ if an oxygen mass fraction of 2\% or 4\% is introduced; 
respectively.

\section{Conclusions}
\textit{Ab initio} simulations showed that, in the diluted limit, ternary mixtures 
made of hydrogen and helium in roughly solar abundance with heavy elements can be 
very well approximated by an isothermal-isobaric additive mixing rule on the 
volume and the energy under thermodynamic conditions relevant in the interior of 
giant planets. It is however necessary to use effective volume and energy for the 
heavy elements because the dominant contributions come from the cross-interactions 
with hydrogen and helium. Although these effective properties tend towards the 
pure species volume and energy at high pressure and temperature, there are 
significant discrepancies at lower pressure and temperature. 

The study of the entropy of oxygen showed that in the diluted limit, the effective 
entropy is mostly influenced by the dissociation of hydrogen but stays rather 
constant along the H-He isentrope in the dissociated regime. If we consider a H-He 
giant planet with a 2\% oxygen mass fraction, the net increase in density is of 
about 1.5\% compared to a pure H-He envelope, for given pressure and temperature. 
Including the entropy correction, the addition of 2\% oxygen increases the 
pressure by roughly 3\% at a given temperature or, equivalently, decreases the 
temperature by less than 2\% at a given pressure. The effect of the entropy is 
thus very small and the net over-density is only of the order of 0.1\%. 

In the case of a two-layer model with an upper layer made of H-He and an oxygen 
enriched inner layer entirely in the metallic regime, the predicted 
pressure-temperature profile do not deviate from the pure H-He predictions. 
Overall, in the diluted limit, the entropy appears to have very little effect on 
the density-pressure relationship.

We argue that the use of the isentrope properties of the H-He mixture with the 
effective volume and energy of the heavy elements as described in this article 
should give a very good approximation of the actual profile of giant planets. The 
comparison with the coming data on the gravitational moments should help to 
constrain the distribution of heavy elements in these planets.

\section*{Acknowledgments}
\acknowledgments

This work was supported by NASA and NSF. Computers at NAS, XSEDE and NCCS were 
used.

\onecolumngrid
\pagebreak

\begin{deluxetable}{cccccc}
\tablecaption{Pressure and internal energy of H-He-$Z$ mixtures with 220 H, 18 He and $N_Z$ heavy entities.\label{table:eosH-He-Z}}
\tablehead{
\colhead{ \multirow{2}{*}{Species}} & \colhead{Temperature} & \colhead{\multirow{2}{*}{$N_Z$}}  & \colhead{Volume} & \colhead{Pressure} & \colhead{Internal Energy} \\
 & \colhead{(K)} &  & \colhead{(\AA$^3$/nucl.)} & \colhead{(GPa)} & \colhead{(eV/nucl.)}
}
\startdata
C &    2000 &4 & 3.88922 &  42.131$\pm$0.040& -2.48961$\pm$0.00099\\
C &    2000 &4 & 3.21772 &  65.040$\pm$0.136& -2.30810$\pm$0.00149\\
C &    2000 &4 & 2.72109 &  95.333$\pm$0.153& -2.09755$\pm$0.00139\\
C &    2000 &4 & 2.36276 & 127.530$\pm$0.345& -1.84461$\pm$0.00370\\
C &    2000 &6 & 6.23309 &  14.638$\pm$0.075& -2.81747$\pm$0.00184\\
C &    2000 &6 & 4.74436 &  27.363$\pm$0.082& -2.68190$\pm$0.00144\\
\enddata
\tablecomments{Table \ref{table:eosH-He-Z} is published in its entirety in the 
electronic edition of the {\it Astrophysical Journal}.  A portion is 
shown here for guidance regarding its form and content.}
\end{deluxetable}
\vspace{-1.2cm}

\begin{deluxetable}{cccccc}
\tablecaption{Pressure and internal energy of H-He-O mixtures with 220 H, 18 He and $N\indice{O}$ O atoms.\label{table:eosH-He-O}}
\tablehead{
 \colhead{Temperature} & \colhead{\multirow{2}{*}{$N\indice{O}$}}  & \colhead{Volume} & \colhead{Pressure} & \colhead{Internal Energy} & \colhead{Helmholtz Free Energy} \\
 \colhead{(K)} & & \colhead{(\AA$^3$/nucl.)} & \colhead{(GPa)} & \colhead{(eV/nucl.)}& \colhead{(eV/nucl.)}
}
\startdata
 5000& 4&  2.27838&  157.151$\pm$0.133& -1.07385$\pm$0.00128& -4.26949$\pm$0.00067 \\
 5000& 4&  1.67448&  317.938$\pm$0.131& -0.43909$\pm$0.00139& -3.42557$\pm$0.00043 \\
 5000& 4&  1.18800&  684.367$\pm$0.170&  0.73063$\pm$0.00165& -2.00307$\pm$0.00031 \\
 5000& 6&  3.19134&   75.024$\pm$0.123& -1.52485$\pm$0.00218& -4.92724$\pm$0.00077 \\
 5000& 6&  2.69878&  109.354$\pm$0.115& -1.33611$\pm$0.00202& -4.64932$\pm$0.00075 \\
 5000& 6&  2.25970&  164.723$\pm$0.129& -1.07571$\pm$0.00102& -4.28233$\pm$0.00057 \\
\enddata
\tablecomments{Table \ref{table:eosH-He-O} is published in its entirety in the 
electronic edition of the {\it Astrophysical Journal}.  A portion is 
shown here for guidance regarding its form and content.}
\end{deluxetable}
\vspace{-1.2cm}

\begin{deluxetable}{cccccc}
\tablecaption{Effective properties of heavy elements in H-He mixtures.\label{table:effprop}}
\tablehead{
 \colhead{\multirow{2}{*}{Species}}& \colhead{Temperature} & \colhead{Pressure}  & \colhead{Volume} & \colhead{Internal Energy} & \colhead{Entropy} \\
&  \colhead{(K)} & \colhead{(GPa)} &\colhead{(\AA$^3$)} &  \colhead{(eV)}& \colhead{($k\indice{B}$)}
}
\startdata
 O &    5000& 1050& 3.018$\pm$0.063& 0.196$\pm$0.718&  8.51$\pm$0.56 \\
 O &    5000& 1100& 2.937$\pm$0.072& 0.281$\pm$0.798&  8.53$\pm$0.61 \\
 O &    5000& 1150& 2.859$\pm$0.081& 0.353$\pm$0.879&  8.55$\pm$0.68 \\
 O &    5000& 1200& 2.784$\pm$0.090& 0.413$\pm$0.962&  8.57$\pm$0.75 \\
 O &    6000&  100& 6.377$\pm$0.082& 5.136$\pm$0.210& \\   
 O &    6000&  150& 5.853$\pm$0.061& 4.740$\pm$0.237& \\   
\enddata
\tablecomments{The effective entropy is only available for oxygen and only at 5000 and 10000~K. Table \ref{table:effprop} is published in its entirety in the 
electronic edition of the {\it Astrophysical Journal}.  A portion is 
shown here for guidance regarding its form and content.}
\end{deluxetable}

\end{document}